\documentclass[11pt]{article}
\usepackage{amsmath}
\allowdisplaybreaks[4]
\usepackage{amssymb}
\usepackage{authblk}
\usepackage{subfigure}
\usepackage{graphicx}
\usepackage{mathrsfs}
\usepackage{tikz}
\usepackage{longtable}
\usepackage{booktabs}
\usepackage{amsthm}
\usepackage[hmargin=2.5cm, vmargin={2.5cm, 3.5cm}]{geometry}
\usepackage{indentfirst}
\usepackage[normalem]{ulem}
\usepackage{natbib}
\usepackage{comment}
\usepackage{setspace}
\usepackage[shortlabels]{enumitem}
\usepackage{ulem}   
\usepackage{color,soul} 
\usepackage{mleftright}

\RequirePackage[colorlinks,citecolor=blue,urlcolor=blue]{hyperref}

\title{Investigating an Alternative for Estimation from a Nonprobability Sample: Matching plus Calibration}
\author[1]{Zhan Liu}
\author[2]{Richard Valliant}
\affil[1]{School of Mathematics and Statistics, Hubei University, China; eleen\_20040109@163.com}
\affil[2]{Research Prof. Emeritus, Universities of Michigan \& Maryland, USA; valliant@umich.edu}

\expandafter\def\csname @fnsymbol\endcsname#1{\ifcase#1\or*\else\fi}

\numberwithin{equation}{section}

\begin{document}
\pagenumbering{arabic}

\maketitle
\textbf{Abstract}: Matching a nonprobability sample to a probability sample is one strategy both for selecting the nonprobability units and for
weighting them. This approach has been employed in the past to select subsamples of persons from a large panel of volunteers. One method of
weighting, introduced here, is to assign a unit in the nonprobability sample the weight from its matched case in the probability sample. The
properties of resulting estimators depend on whether the probability sample weights are inverses of selection probabilities or are calibrated.
In addition, imperfect matching can cause estimates from the matched sample to be biased so that its weights need to be adjusted, especially
when the size of the volunteer panel is small. Calibration weighting combined with matching is one approach to correcting bias and reducing
variances. We explore the theoretical properties of the matched and matched, calibrated estimators with respect to a quasi-randomization
distribution that is assumed to describe how units in the nonprobability sample are observed, a superpopulation model for analysis variables
collected in the nonprobability sample, and the randomization distribution for the probability sample. Numerical studies using simulated and
real data from the 2015 US Behavioral Risk Factor Surveillance Survey are conducted to examine the performance of the alternative estimators.
\doublespacing

\pagebreak

\section{Introduction} \label{sec:intro}
Probability samples have been the standard for finite population estimation for many decades. However, probability samples can have many
nonsampling problems like low contact and response rates or missing data for units that do respond. Response rates in US surveys, in
particular, have been declining for at least two decades \citep{Brick.2013}. Since nonprobability samples can be faster and cheaper to
administer and collect, some organizations are gravitating toward their use \citep{Terhanian.2012}. \citet{BakerBrick.2013} review the reasons
that nonprobability samples, like volunteer internet panels, may be used rather than a probability sample. Among them are lower costs and
compressed data collection periods. Quick turnaround can be especially important to gauge public well-being in health crises like the COVID-19
pandemic of 2020.

There are a variety of problems with nonprobability samples, especially among persons in a panel that have been recruited to
participate in future surveys \citep[e.g., see][]{BakerBrick.2013,Valliant.2011,VDK.2018}. These include selection bias, coverage error, panel
nonresponse, attrition, and measurement error. We concentrate on the use of matching and calibration to adjust for the first two of
these---selection bias and coverage error. Selection bias occurs if the sample differs from the nonsample in such a way that the sample cannot
be projected to the full population without some type of statistical adjustment. Coverage error can occur if, for example, a volunteer panel
consists of only persons with access to the Internet, assuming that the entire population of a country is the target of the survey. Other, more
subtle forms of coverage error can occur if certain demographic groups would rarely or never participate in the particular type of
nonprobability survey being conducted.

Because the selection of a nonprobability sample is not controlled by a survey designer, estimation methods other than standard design-based
approaches are needed. At least six alternatives can be considered for weighting and estimation with nonprobability samples:
\begin{enumerate} [(1)]
    \item Na\"ive method where all units are assigned the same weight
    \item Quasi-randomization where a pseudo-inclusion probability is estimated for each nonprobability unit
    \item Superpopulation modeling of analysis variables ($Y$'s)
    \item Doubly robust estimation where quasi-randomization and superpopulation modeling are combined
    \item Mass imputation of $Y$'s into a probability sample using values from a nonprobability sample to form an imputation model
    \item Matching of a nonprobability sample to a probability sample whose units are used as donors of weights to the nonprobability sample
\end{enumerate}

The na\"ive method of equal weighting is rarely, if ever, appropriate because nonprobability samples are not generally distributed
proportionally across demographic or other important groups in the population. Alternatives (2)-(4) were reviewed by \citet{Elliott.2017} and
\citet{Valliant.2020} and further studied by \cite{CLW.2020}. \citet{Wang.2020b} refined alternative (2) by kernel-smoothing the propensity
weights. Alternative (5) was proposed by \citet{Kim.2021} and involves fitting an imputation model using data from the nonprobability sample
and imputing $Y$ values to the units in the probability sample using that model. The probability sample with imputed values is provided to
analysts but not the nonprobability sample. Mass imputation solves the weighting problem by using the weights associated with the probability
sample. The dissertation of \citet{Wang.2020a} studied a version of (6) in which a kernel-smoothing method was used to proportionally
distribute the probability sample weights to units in the nonprobability sample. We study another, somewhat simpler version of alternative (6),
and particularly address some problems with the method.

\subsection{Notation and Models Used for Analysis} \label{subsec:notation}

Both a probability sample, denoted by $S_p$ and a nonprobability sample, denoted by $S_{np}$ will be used in subsequent sections. The target
population for which estimates are made is $U$ and has $N$ units. To examine properties of estimators, three distributions will be used.
Expectations taken with respect to the sample design used to select the probability sample $S_p$ will be denoted by a $\pi$ subscript. The
probability of selection of unit $i$ in $S_p$ is $\pi_i$. To analyze the nonprobability sample $S_{np}$, we assume that its units are selected
by an unknown quasi-randomization distribution; expectations taken with respect to that distribution will be specified by a subscript $R$. The
probability that unit $j$ is included in $S_{np}$ is $R(\mathbf{x}_j)$ where $\mathbf{x}_j$ is a $C$-vector of covariates or auxiliaries
associated with unit $j$. To simplify notation in later sections, we set $R(\mathbf{x}_j) \equiv R_j$. The analysis variable $Y$ will also be
assumed to be generated by a superpopulation model, $\xi$. Consider the linear model for $Y_i$ defined by
\begin{equation} \label{eq:model}
    Y_i = \mathbf{x}_i \boldsymbol{\beta} + \epsilon_i\ (i \in U),
\end{equation}
where $\boldsymbol{\beta}$ is a $C \times 1$ parameter vector, and the $\epsilon_i$ are independent, random errors with mean zero and variance
$\sigma_i^2$. Theory for nonlinear models can also be worked out, as in \citet{CLW.2020}, but a linear model is convenient for purposes here.
Under model \eqref{eq:model}, the expected value of the population total, $Y_U=\sum_{i \in U} Y_i$, is $E_{\xi}\left(Y_U \right)
=\mathbf{X}_U\boldsymbol{\beta}$ where $\mathbf{X}_U=\sum_{i \in U} \mathbf{x}_i$.

The remainder of the article is organized as follows. Section \ref{sec:Apps} describes how matching can be applied to obtain basic weights for
units in the nonprobability sample and reviews the methods of matching.  Section \ref{sec:estmatch} presents the theory for bias and variance in
different situations. Section \ref{sec:calib} investigates properties of matched estimators when the nonprobability sample is calibrated to
population totals of covariates. In Section \ref{sec:sim}, the sample matching and the calibration adjustment are applied in a simulation study
using artificial data. In Section \ref{sec:realdata}, an application to a real population is conducted to evaluate the performance of the
proposed estimates. The last section summarizes our findings.

\section{Applications of Matching} \label{sec:Apps}
Sample matching has been an option for estimating treatment effects in causal inference for some time \citep[e.g.,
see][]{Cochran.1953,Rubin.1973,Rosenbaum.1983}. Moreover, it has been widely applied in evaluation research, observational studies and
epidemiological studies \citep{Rothman.2008}. More recently, it also has been applied as a way of identifying a sample in market research,
public opinion surveys \citep[e.g.,][]{Vavreck.2008,Terhanian.2012}, and other nonprobability sampling surveys, especially using volunteer
panel surveys. \citet{BakerBrick.2013} review some of the applications of matching in survey sampling. Its purpose in nonprobability sampling
surveys is to reduce selection bias and to estimate population characteristics. Another application of statistical matching is to overcome the
problem of missing data created when some persons do not consent to having their survey responses linked to administrative databases
\citep{Gessendorfer.2018}.

The basic idea of sample matching in survey sampling is that first a random, probability sample, $S_p$, is selected from the sampling frame of
the target population. This probability sample is matched to a pool of nonprobability cases, e.g., a volunteer panel of persons. The resulting
matched sample from the nonprobability pool is denoted by $S_{np}$. The probability sample should have none of the coverage problems of the
nonprobability sample. This probability sample is also called a \textit{reference} sample \citep{Lee.2006} and can be an existing survey (or
subsample of one) rather than one specially conducted to serve as the reference. For example, in the US the American Community Survey (ACS,
\url{https://www.census.gov/programs-surveys/acs}) is one possibility for a large, well-conducted household, reference survey. The probability
sample should be representative of the target population in the sense that it can be used to make unbiased and/or consistent estimates of
population quantities. We assume that $S_p$ does not include the $Y$ variables for which estimates are to be made; these are collected from the
nonprobability sample.

The application of matching described by \citet{Rivers.2007} is one in which $S_p$ is a simple random sample (\textit{srs}). The nonprobability
sample $S_{np}$ is obtained by a one-to-one match of $S_p$ to a much larger pool of nonprobability cases, yielding a set $S_{np}$ of the same
size as $S_p$.  Since $S_{p}$ was treated as an \textit{srs}, every unit in $S_{np}$ was given the same weight. When $S_{p}$ is an
\textit{srs}, the distribution across various characteristics of $S_{np}$ is expected to be the same as that of the population. However, in an
evaluation of the nonprobability samples offered by nine commercial vendors, \citet{Kennedy.2016} found that a nonprobability sample may still
produce biased estimators even though it had the same demographic distribution as the population. In other words, matching to an \textit{srs}
$S_p$ to obtain $S_{np}$ can be inadequate without further weighting. \citet{Rivers.2009} describe an election polling application where the
sample was obtained by matching, as described above, but inverses of estimated propensities of being in the nonprobability sample were used as
weights.

Sample matching in alternative (6), as applied in this paper, fits into the quasi-randomization approach. Each unit in a probability sample is
matched to a unit in the nonprobability sample based on a set of covariates. The logical extension of \citet{Rivers.2007} is for the
probability sample unit to ``donate'' its weight to the matched, nonprobability sample unit. The intuitive argument to justify this is that if
the nonprobability units match the probability units on an extensive list of covariates, then the $S_{np}$ units are exchangeable for the $S_p$
units, $S_{np}$ constitutes the same sort of sample as $S_p$, and the units in $S_{np}$ can be weighted in the same way. This approach has the
advantage of straightforward retention of all analytic data collected in the nonprobability sample unlike alternative (5) which could require a
separate imputation model for every $Y$ variable.

The probability sample used for matching can be larger, smaller, or equal in size to the nonprobability sample, although the method in which
$S_{np}$ is selected to have the same size as $S_p$ has advantages. If a pool of nonprobability units is used that is much larger than the
probability sample, finding a close match for each unit in the probability sample may be more feasible. This would be the case when a large
panel of volunteers has been accumulated. If the nonprobability sample is smaller, a unit in $S_{np}$ may be matched to more than one unit in
$S_p$, making it unclear how to weight the $S_{np}$ cases. In this article, we assume that the resulting sample size of the matched,
nonprobability sample, $S_{np}$, equals the sample size of the probability sample, $S_{p}$, that it is matched against. Denote this sample size
by $n$.

\subsection{Methods of Matching}
Which matching algorithm to use is a question. There are various algorithms, including nearest neighbour matching, caliper and radius matching,
stratification and interval matching, as well as kernel and local linear matching \citep{Caliendo.2008}. Among these matching algorithms,
nearest neighbour matching based on Euclidean distance is most straightforward. It contains, as special cases, single nearest neighbour
matching without replacement, single nearest neighbour matching with replacement and multiple nearest neighbour matching. In single nearest
neighbour matching, for a unit in $S_p$ only one unit from the nonprobability pool can be chosen as its matching unit based on the covariates
present in both datasets. If single nearest neighbour matching is done without replacement, a unit in the nonprobability pool can be chosen
only once as a match. Single nearest neighbour matching without replacement may, however, have poor performance when the target sample and the
volunteer panel have very different covariate distributions \citep{Dehejia.2002}. To overcome this problem, single nearest neighbour matching
with replacement and multiple nearest neighbour matching were proposed to increase the average quality of matching and reduce the bias
\citep{Smith.2005}. Other matching methods have been suggested that use more than one unit in the nonprobability pool as the matching unit for
an individual in the probability sample. Caliper and radius matching use this approach.

Another issue in sample matching is that the matching process will become relatively more difficult as the number of relevant covariates
increases. This is the curse of dimensionality noted by \citet{Rosenbaum.1983}. In order to solve this problem, they propose the propensity
score, which is the conditional probability of receiving a treatment given the covariates $X$, denoted by $p(X)=P(D=1|X)$, where $D$ is the
binary indicator taking either the value 1 (receiving treatment, e.g. participation in a volunteer panel) or 0 (not receiving treatment).
\citet{Rosenbaum.1983} have proved that matching on the propensity score $p(X)$ is also valid when it is valid to match on the covariates $X$.
Compared with direct matching based on all covariates, propensity score matching can reduce multiple dimensions (many covariates) to a single
dimension, greatly simplifying the matching process.  Consequently, it has been widely used in medical and epidemiological studies, economics,
market research and a host of other fields \citep{Schonlau.2009, BakerBrick.2013}.

\section{Estimation from Matched Sample} \label{sec:estmatch}

In this section, we introduce estimators of means and totals based on the matched sample $S_{np}$ and derive their properties. An estimator of
a population total is
\begin{equation} \label{eq:est.tot}
    \widehat{Y}_M = \sum_{j \in S_{np}} \widetilde{w}_j y_j
\end{equation}
where $\widetilde{w}_j$ is the weight from the probability sample unit that is matched to unit $j$ in $S_{np}$ and $y_j$ is the $Y$ value
observed for that unit. We assume that these weights are appropriately scaled for estimating population totals. In particular, $\hat{N}_M =
\sum_{j \in S_{np}} \widetilde{w}_j$ is an estimator of $N$, the size of the target population. The mean of $Y$ is estimated by
$\widehat{\overline{Y}}_M = \widehat{Y}_M / \widehat{N}_M$. We will consider two cases of weighting of the probability sample $S_p$:
\begin{enumerate}[(i)]
    \item The weight used for each unit in $S_p$ is the inverse of the selection probability of that unit, i.e., $\widetilde{w}_j =
        \pi_j^{-1}$; the estimator of the total with this weight is denoted by $\hat{Y}_{M1}$ subsequently;
    \item The $S_p$ weights are those for a general regression (GREG) estimator; the estimator with this weight is denoted by $\hat{Y}_{M2}$.
\end{enumerate}
Note that the GREG in case (ii) includes the commonly used poststratification estimator. Whether those weights are related to the
pseudo-inclusion probabilities of the units in $S_{np}$ largely determines whether $\widehat{Y}_{M1}$ and $\widehat{Y}_{M2}$ are biased or not
as shown below. The arguments given are largely heuristic, although they can be formalized using technical conditions like those in
\citet{CLW.2020}.

Properties of estimators can be calculated in several ways: with respect to the $\xi$-model only, with respect to the $R$-distribution only,
with respect to the $\pi$-distribution, or with respect to a combination of the distributions. In subsequent sections, we compute biases and
variances with respect to the combined $R \pi$-distribution. The $R \pi$ calculation is analogous to the design-based calculations used in much
of sampling theory. In addition, bias and variance calculations are made with respect to the $\xi$-model and combined $R\pi\xi$-distributions.
The calculations made with respect to the $\xi$-distribution are conditional on the $S_{np}$ and $S_p$ samples. In principle, $\xi$
calculations are more reflective of the statistical properties for the particular sets of units in $S_{np}$ and $S_p$.

\subsection{Bias of the matched estimator for case (i)} \label{subsec:bias.c1}

Taking the expectation of $\widehat{Y}_{M1} - Y_U$ under case (i) with respect to the pseudo-randomization distribution only gives
\begin{align*}
    E_R\left(\widehat{Y}_{M1} - Y_U \right) &= E_R\left( \sum_{j \in S_{np}} \pi_{j}^{-1} y_j \right) - Y_U \\
        &= \sum_{j \in U} \frac{R_j}{\pi_{j}} y_j- Y_U\,,
\end{align*}
If $R_j = \pi_j$, then $\widehat{Y}_M$ will be $R$-unbiased. However, this does not have to be true generally. For example, if $R_j = Pr\left(j
\in S_{np} \mid \mathbf{x}_j\right)$ is a complicated logistic function of a set of covariates that were not used in determining
$\widetilde{w}_j$, the estimator is $R$-biased. Another situation leading to $R$-bias would be when the pseudo-inclusion mechanism is
\textit{nonignorable}, i.e., $Pr\left(j \in S_{np} \mid \mathbf{x}_j, y_j\right) \neq Pr\left(j \in S_{np} \mid \mathbf{x}_j\right)$. Since in
a probability sample, the selection mechanism is always ignorable, $\widetilde{w}_j \neq 1/Pr\left(j \in S_{np} \mid \mathbf{x}_j, y_j\right)$
when inclusion in the nonprobability sample depends on $Y$.

If the expectation is taken over the $Y$-model, the result is
\begin{align*}
     E_{\xi}\left(\widehat{Y}_{M1} - Y_U \middle\vert S_{np}, S_p\right) &= \left( \mathbf{\widehat{X}}_{np}(\pi) - \mathbf{X}_U \right)
     \boldsymbol{\beta}
\end{align*}
where $\widehat{X}_{np}(\pi) = \sum_{S_{np}} \mathbf{x}_j/\pi_j$. The $\xi$-bias is non-zero unless $\mathbf{\widehat{X}}_{np}(\pi) =
\mathbf{X}_U$. If $\widehat{\mathbf{X}}_{np}(\pi)$ is an unbiased estimator of $\mathbf{X}_U$ under the quasi-randomization $R$-distribution,
$\widehat{Y}_{M1}$ will be unbiased when averaged over both the $R$- and $\xi$-distributions (and, equivalently, over the $R$, $\pi$, and $\xi$
distributions). But, as for $\widehat{Y}_{M1}$, $\widehat{\mathbf{X}}_{np}(\pi)$ will be biased if the correct $R$-model is not linked to the
$S_p$ weights, i.e., if $\widetilde{w}_j = {\pi}_j^{-1} \neq 1/R_j$.

\subsection{Bias of the matched estimator for case (ii)} \label{subsec:bias.c2}
If the weights from the probability sample have been calibrated to population totals of some covariates $\mathbf{x}$, the bias calculation
changes somewhat. Take the case of the general regression (GREG) estimator being used for $S_p$. That is, $\widetilde{w}_j = g_j/\pi_j$ where
\begin{equation} \label{eq:gj}
    g_j = 1 + \left(\mathbf{X}_U - \widehat{\mathbf{X}}_{p} \right)^T \widetilde{\mathbf{A}}_{p}^{-1} \mathbf{x}_j/\widetilde{\sigma}_j^2
\end{equation}
with $\widehat{\mathbf{X}}_{p} = \sum_{S_{p}} \mathbf{x}_j/\pi_j$ and $\widetilde{\mathbf{A}}_p =\sum_{S_{p}} \mathbf{x}_j \mathbf{x}_j^T \big/
\left(\pi_j \widetilde{\sigma}_j^2 \right)$.  The values of $\widetilde{\sigma}_j^2$ are often set to a constant in practice, but for
completeness, we include $\widetilde{\sigma}_j^2$ in subsequent formulas. If $\widetilde{\sigma}_j^2$'s are used in estimators of totals, they
will be generally be assumed values of the model variances in \eqref{eq:model}; but, we do not require that $\widetilde{\sigma}_j^2 =
\sigma_j^2$. Note also that the $\pi_j$'s must be available separately for each unit in the probability sample in order to recover
$\mathbf{A}_{p}$ separately from the $\widetilde{w}_j$. In some public-use files, users may only be presented with the $\widetilde{w}_j =
g_j/\pi_j$ and not $\pi_j$.

The estimator of the total is then
\begin{align} \label{eq:YM.greg}
    \widehat{Y}_{M2} & = \widehat{Y}_{np}(\pi) + \left(\mathbf{X}_U - \widehat{\mathbf{X}}_{p} \right)^T \widetilde{\mathbf{A}}_{p}^{-1}
    \sum_{j \in S_{np}} \mathbf{x}_j y_j/ (\pi_j \widetilde{\sigma}_j^2)
\end{align}
where $\widehat{Y}_{np}(\pi) = \sum_{j \in S_{np}} y_j/\pi_j$. The $\xi$-bias is
\begin{equation} \label{eq:YM.xi.bias}
    E_{\xi}\left(\widehat{Y}_{M2} - Y_U \middle\vert S_{np}, S_p \right) = \widehat{\mathbf{X}}_{np}(\pi) \boldsymbol{\beta} +
    \left(\mathbf{X}_U - \widehat{\mathbf{X}}_p \right)^T \widetilde{\mathbf{A}}_p^{-1} \widetilde{\mathbf{A}}_{np}(\pi) \boldsymbol{\beta} -
    \mathbf{X}_U \boldsymbol{\beta}
\end{equation}
where $\widetilde{\mathbf{A}}_{np}(\pi) = \sum_{S_{np}} \mathbf{x}_j \mathbf{x}_j^T \big/ \left(\pi_j \widetilde{\sigma}_j^2 \right)$. Thus,
$\widehat{Y}_{M2}$ is $\xi$-model biased even though the weights in $S_p$ are calibrated on the $\mathbf{x}$'s. The $R$-expectation (which is
also the $R\pi$-expectation) is
\[ E_R \left(\widehat{Y}_{M2} \right) = \sum_U \frac{R_j}{\pi_j} y_j + \left(\mathbf{X}_U - \widehat{\mathbf{X}}_{p} \right)^T
\widetilde{\mathbf{A}}_{p}^{-1} \sum_U \frac{R_j}{\pi_j} \frac{\mathbf{x}_j y_j}{\widetilde{\sigma}_j^2} \]
which is also generally not equal to $Y_U$.

If $R_j = \pi_j$ and sampling for $S_{np}$ and $S_p$ is ignorable, reasonable assumptions are that $N\widetilde{\mathbf{A}}_{p}^{-1}$ and
$N^{-1}\widetilde{\mathbf{A}}_{np}(\pi)$ converge in probability to $N^{-1}\widetilde{\mathbf{A}}_U = N^{-1}\sum_U \mathbf{x}_j \mathbf{x}_j^T
\big/ \widetilde{\sigma}_j^2$. (See assumption (v) in the Appendix). In that case, $\widetilde{\mathbf{A}}_p^{-1}
\widetilde{\mathbf{A}}_{np}(\pi) \overset{p}{\to} \mathbf{I}_C$ with $\mathbf{I}_C$ being the $C \times C$ identity matrix, and
$E_{\xi}\left(\widehat{Y}_{M2} - Y_U \middle\vert S_{np}, S_p \right) \rightarrow \left( \widehat{\mathbf{X}}_{np}(\pi) -
\widehat{\mathbf{X}}_p \right)\boldsymbol{\beta}$.

Taking the expectation of \eqref{eq:YM.xi.bias} with respect to the $R$- and $\pi$-distributions shows that $\widehat{Y}_{M2}$ is approximately
$R\pi\xi$-unbiased, but this depends on $R_j = \pi_j$ for all units in $S_{np}$. Under the same conditions (i.e., $R_j = \pi_j$ and
$N\widetilde{\mathbf{A}}_p^{-1}$ and $N^{-1}\widetilde{\mathbf{A}}_{np}(\pi)$ converging),
\[ E_R \left(\widehat{Y}_{M2} \right) \doteq Y_U + \left(\mathbf{X}_U - \widehat{\mathbf{X}}_{p} \right)\widetilde{\mathbf{B}}_U\]
where $\widetilde{\mathbf{B}}_U = \widetilde{\mathbf{A}}_U^{-1} \sum_U \mathbf{x}_j y_j \big/ \widetilde{\sigma}_j^2$. Consequently, $E_R
E_{\pi} \left(\widehat{Y}_{M2} \right) \doteq Y_U$, assuming that $\widehat{\mathbf{X}}_p$ is $\pi$-unbiased. Similarly, $E_R E_{\pi} E_\xi
\left(\widehat{Y}_{M2} - Y_U\right) = 0$.

The results in sections \ref{subsec:bias.c1} and \ref{subsec:bias.c2} can be summarized as follows:
\begin{itemize}
    \item Case (i), $\widetilde{w}_j$ is the inverse of the selection probability for its matched unit in the probability sample, $S_p$,
        i.e., $\widetilde{w}_j = \pi_j^{-1}$
    \begin{itemize}
        \item $\widehat{Y}_{M1}$ is $\xi$-biased when the linear model \eqref{eq:model} holds;
        \item $\widehat{Y}_{M1}$ is $R\pi$-unbiased if $R_j = \pi_j$, i.e., the probability of a unit's being in the nonprobability sample,
            $S_{np}$, equals its probability of being in the probability sample, $S_p$;
        \item $\widehat{Y}_{M1}$ is $R\pi\xi$-unbiased when the linear model \eqref{eq:model} holds and $R_j = \pi_j$;
    \end{itemize}
    \item Case (ii), $\widetilde{w}_j$ is the GREG weight for its matched unit in $S_p$
    \begin{itemize}
        \item $\widehat{Y}_{M2}$ is $\xi$-biased under \eqref{eq:model} even though $S_p$ is calibrated on the $x$'s in the model;
        \item $\widehat{Y}_{M2}$ is $R$-biased in general;
        \item $\widehat{Y}_{M2}$ is approximately $R\pi$-unbiased and $R\pi\xi$-unbiased if $R_j = \pi_j$;
    \end{itemize}
\end{itemize}
The key requirement (in addition to ignorability) for unbiasedness of any type is that the observation probability of a unit in the
nonprobability sample should be equal to the selection probability of its matched unit from the probability sample. This seems unlikely to be exactly true in most applications.

\subsection{Variance of the Matched Estimator in case (i)} \label{subsec:Vcase.i}
Since a variance estimator is mainly useful in a situation where a point estimator is unbiased or consistent, we concentrate on the case where
$R_j = \pi_j$ and $\widehat{Y}_{M1}$ is $R$-unbiased. Calculation of the variance of $\widehat{Y}_{M1}$ with respect to the pseudo-inclusion
probability distribution depends on the joint distribution of the indicators, $\{\delta_j\}_{j \in U}$ where $\delta_j = 1$ if $j \in S_{np}$
and 0 if not. If the $\delta_j$ have the same joint distribution as that of the indicators for being in the probability sample, then $S_{np}$
can be treated as having the same sample design as $S_p$. If so, $V_R\left( \widehat{Y}_{M1} \right) = V_{\pi}\left( \widehat{Y}_{M1} \right)$,
and the variance estimator for $\widehat{Y}_{M1}$ would be determined by the sample design for $S_p$. For example, if the probability sample
was a stratified, cluster sample, then the variance estimator appropriate to that design would be used.

When $\widetilde{w}_{j} = \pi_j^{-1}$, a more realistic assumption, given the way that nonprobability samples are often acquired, is to treat
the $\{\delta_j\}_{j \in U}$ as being independent. With that assumption, the $R$-variance can be estimated with a formula appropriate for a
Poisson sample. Another option is the formula for a sample selected with replacement and with probabilities equal to $R_j = \pi_j$:
\begin{align} \label{eq:VR.hat}
    v_{R\pi}\left(\widehat{Y}_{M1} \right)
        &= \frac{n}{n-1} \sum_{j \in S_{np}} \left(\widetilde{w}_{j}y_{j} - \frac{1}{n} \sum_{j^{\prime} \in S_{np}}
        \widetilde{w}_{j^\prime}y_{j\prime}\right)^2\,.
\end{align}
Because $\widehat{Y}_{M1}$ does not depend on $S_p$, $V_{R \pi}(\widehat{Y}_{M1}) = V_{R}(\widehat{Y}_{M1})$ and \eqref{eq:VR.hat} can be
interpreted as an estimator of either.  Estimator \eqref{eq:VR.hat} is convenient because it is the default in survey software packages like R
\texttt{survey}, Stata, and SAS. However, as shown in Appendix \ref{app:vRexp}, $v_{R\pi}$ is a biased estimator of the model variance given
below.

The $\xi$-variance under \eqref{eq:model} in case (i) is $V_{\xi}\left(\widehat{Y}_{M1} \right) = \left(\sum_{S_{np}}\sigma_j^2 \big/ \pi_j^2
\right)$ which can be estimated by
\begin{equation} \label{eq:VxM1c1}
    v_{\xi}\left(\widehat{Y}_{M1} \right) = \sum_{S_{np}} \frac{e_j^2 }{\pi_j^2}\,,
\end{equation}
where $e_j^2 = \left(y_j - \mathbf{x}_j^T \widehat{\widetilde{\mathbf{B}}}_{np}(\pi)\right)^2$ is an approximately $\xi$-unbiased estimator of
$\sigma_j^2$ with
\[
    \widehat{\widetilde{\mathbf{B}}}_{np}(\pi) = \left(\sum_{S_{np}} \mathbf{x}_j \mathbf{x}_j^T / (\pi_j \widetilde{\sigma}_j^2)
    \right)^{-1} \sum_{S_{np}} \mathbf{x}_j y_j / (\pi_j \widetilde{\sigma}_j^2)
\]
\citep{MacKinnon.1985}. Note that, because the $Y$'s are not available in the probability sample, we must estimate $\boldsymbol{\beta}$ from
the nonprobability sample.

The $R\pi\xi$-variance, in general, is equal to
\begin{align} \label{eq:3partvar}
    V_{R\pi\xi}\left( \widehat{Y}_{M1} \mid S_p,S_{np}\right) &= E_R \left\{V_{\pi\xi}\left( \widehat{Y}_{M1} \mid S_p,S_{np}\right) \right\} +
        V_R \left\{E_{\pi\xi}\left( \widehat{Y}_{M1} \mid S_p,S_{np} \right) \right\} \notag \\
    &= E_R \left\{ E_{\pi}V_{\xi}\left(\widehat{Y}_{M1} \mid S_p,S_{np}\right) + V_{\pi}E_{\xi}\left(\widehat{Y}_{M1} \mid S_p,S_{np}\right)
    \right\} + \notag \\
        & \qquad V_R \left\{E_{\pi} E_{\xi}\left(\widehat{Y}_{M1} \mid S_p,S_{np}\right) \right\}\,.
\end{align}
For case (i) $\widehat{Y}_{M1}$ does not depend on $S_p$ and \eqref{eq:3partvar} reduces to $V_{R\pi\xi}\left( \widehat{Y}_{M1} \mid
S_p,S_{np}\right) = V_{R\xi}\left( \widehat{Y}_{M1} \mid S_p,S_{np}\right) = E_R \left\{V_{\xi}\left( \widehat{Y}_{M1} \mid S_p,S_{np}\right)
\right\} +
        V_R \left\{E_{\xi}\left( \widehat{Y}_{M1} \mid S_p,S_{np} \right) \right\} $.

As shown in Appendix \ref{app:vRexp}, for case (i) with $R_j = \pi_j$, this is
\begin{equation} \label{eq:YM1.VRpi}
    V_{R\pi \xi}\left(\widehat{Y}_{M1} \right) = \sum_U \frac{\sigma_j^2}{\pi_j} + \boldsymbol{\beta}^T
    V_R(\widehat{\mathbf{X}}_{np})\boldsymbol{\beta}\,.
\end{equation}
Notice that, even though $\hat{Y}_{M1}$ does not directly depend on $x$, the $R\pi\xi$-variance does after accounting for the $\xi$-model
structure. Expression \eqref{eq:YM1.VRpi} can be estimated by
\begin{equation} \label{eq:vRpixi.M1c1}
     v_{R\pi\xi}\left(\widehat{Y}_{M1} \right) = \sum_{S_{np}} \frac{e_j^2}{\pi_j^2} + \widehat{\widetilde{\mathbf{B}}}_{np}(\pi)^T
     v_R(\widehat{\mathbf{X}}_{np})\widehat{\widetilde{\mathbf{B}}}_{np}(\pi)
\end{equation}
where $v_R(\widehat{\mathbf{X}}_{np})$ is, for example, a version of \eqref{eq:VR.hat} adapted to estimate a covariance matrix.

\subsection{Variance of the Matched Estimator in case (ii)} \label{subsec::Vcase.ii}
The $\xi$-model variance is $V_{\xi}\left( \widehat{Y}_{M2} \right) = \sum_{S_{np}} (g_j/\pi_j)^2 \sigma_j^2$, which can be estimated by
\begin{equation} \label{eq:vxi.M2c2}
    v_{\xi}\left( \widehat{Y}_{M2} \right) = \sum_{S_{np}} (g_j/\pi_j)^2 e_j^2.
\end{equation}
As noted in Appendix \ref{app:YM2c2var}, the estimator of total can be approximated by
\begin{equation} \label{eq:YM2approx}
    \widehat{Y}_{M2} \doteq \widehat{Y}_{np}(\pi) + \left(\mathbf{X}_U - \widehat{\mathbf{X}}_p  \right) \tilde{\mathbf{B}}_U
\end{equation}
and the approximate $R\pi$ variance is
\begin{equation} \label{eq:VRpi.M2c2}
    V_{R\pi}\left( \widehat{Y}_{M2} \right) \doteq V_R\left(\widehat{Y}_{np}(\pi) \right) + \tilde{\mathbf{B}}_U^T
    V_{\pi}(\widehat{\mathbf{X}}_p) \tilde{\mathbf{B}}_U \,,
\end{equation}
which can be estimated as
\begin{equation} \label{eq:vRpi.M2c2}
    v_{R\pi}\left( \widehat{Y}_{M2} \right) = v_R\left(\widehat{Y}_{np}(\pi) \right) + \widehat{\widetilde{\mathbf{B}}}_{np}(\pi)^T
    v_{\pi}(\widehat{\mathbf{X}}_p) \widehat{\widetilde{\mathbf{B}}}_{np}(\pi).
\end{equation}

Note that, in the situation studied here, both terms of the variance in \eqref{eq:VRpi.M2c2} have the same order of magnitude, $O(N^2/n)$,
since they are based on samples of the same size. Thus, the ${R\pi}$-variance is the variance in the nonprobability sample of the estimator
with inverse pseudo-inclusion probability weights plus a term reflecting the variance of the estimator of the \textbf{x}-totals in the
probability sample.

Since $\widehat{Y}_{np}(\pi) = \widehat{Y}_{M1}$, \eqref{eq:VRpi.M2c2} also implies that the $R\pi$-variance of $\widehat{Y}_{M2}$ with
calibrated $S_p$ weights is larger than that of the uncalibrated $\widehat{Y}_{M1}$. This disagrees with the usual expectation that calibration
on an effective predictor of $Y$ reduces variances. To better understand this, note that if the matched $x$'s in $S_p$ and $S_{np}$ were
identical, then $\widehat{\mathbf{X}}_p = \widehat{\mathbf{X}}_{np}$ and the variable part of \eqref{eq:YM2approx} could be written as a
weighted sum over $S_{np}$ of residuals, which can then be used to show that $\widehat{Y}_{M2}$ can have a smaller variance than
$\widehat{Y}_{M1}$. However, with imperfect matching the relationship in \eqref{eq:VRpi.M2c2} becomes more realistic.

As shown in Appendix \ref{app:YM2c2var}, the approximate $R\pi\xi$-variance when $R_j=\pi_j$ is
\begin{equation} \label{eq:VRpixi.M2c2}
    V_{R\pi\xi}\left( \widehat{Y}_{M2} \right)
     \doteq \sum_{U} \frac{\sigma_j^2}{\pi_j} + \boldsymbol{\beta}^T V_R \left( \widehat{\mathbf{X}}_{np}\right)\boldsymbol{\beta} +
     \boldsymbol{\beta}^T V_{\pi} \left( \widehat{\mathbf{X}}_{p}\right)\boldsymbol{\beta}\,.
\end{equation}
A natural estimator of \eqref{eq:VRpixi.M2c2} is then
\begin{equation} \label{eq:vRpixi.M2c2}
    v_{R\pi\xi}\left( \widehat{Y}_{M2} \right) = \sum_{S_{np}} (e_j / \pi_j)^2 + \widehat{\widetilde{\mathbf{B}}}_{np}(\pi)^T v_R \left(
    \widehat{\mathbf{X}}_{np}\right) \widehat{\widetilde{\mathbf{B}}}_{np}(\pi) +
    \widehat{\widetilde{\mathbf{B}}}_{np}(\pi)^T v_{\pi} \left( \widehat{\mathbf{X}}_{p}\right) \widehat{\widetilde{\mathbf{B}}}_{np}(\pi).
\end{equation}

Consequently, there are several options for variance estimation for $\widehat{Y}_{M2}$ for cases (i) and (ii). They can be summarized as:
\begin{itemize}
    \item Case (i), $\widetilde{w}_j = \pi_j^{-1}$
    \begin{itemize}
        \item Estimate the $\xi$-variance with \eqref{eq:VxM1c1}
        \item Estimate the quasi-randomization ($R\pi$) variance with the with-replacement estimator in \eqref{eq:VR.hat};
        \item Estimate the $R\pi\xi$-model variance with $v_{R\pi\xi}$ in \eqref{eq:vRpixi.M1c1};
    \end{itemize}
    \item Case (ii), $\widetilde{w}_j$ = GREG weight from $S_p$
    \begin{itemize}
        \item Estimate the $\xi$-variance with \eqref{eq:vxi.M2c2}
        \item Estimate the $R\pi$-variance with \eqref{eq:vRpi.M2c2};
        \item Estimate the $R\pi\xi$-variance with \eqref{eq:vRpixi.M2c2};
    \end{itemize}
\end{itemize}

\section{Calibration Adjustment After Matching} \label{sec:calib}
The $R-$, $R\pi-$, or $R\pi\xi$-bias of the matched estimators, $\widehat{Y}_{M1}$ and $\widehat{Y}_{M2}$, in section \ref{sec:estmatch} depend
critically on whether $Pr\left(j \in S_{np} \mid \mathbf{x}_j \right) = Pr\left(i \in S_p \right)$ for matched units $i$ and $j$. Matching on
covariates attempts to ensure this; however, there is no guarantee that the condition is satisfied regardless of how extensive the set of
covariates is.

Consequently, one might hope that calibrating the weights for the nonprobability sample will provide some bias protection. Suppose that the
$\{\widetilde{w}_j\}_{j \in S_{np}}$ weights are calibrated to the $\mathbf{X}_U$ population totals using the chi-square distance function
associated with a GREG. Using the standard formula from \citet[][, eqn. 6.5.10]{Sarndal1992}, the resulting weight for unit $j$ is
\begin{align} \label{eq:wjstar}
    w_j^{\ast} &= \widetilde{w}_j \left[1 + \left(\mathbf{X}_U - \widehat{\mathbf{X}}_{np}(\widetilde{w}) \right)^T
    \left[\mathbf{A}_{np}^{\ast}(\widetilde{w})\right]^{-1} \mathbf{x}_j / \sigma_j^{\ast 2} \right] \notag \\
    & \equiv \widetilde{w}_j g_j^{\ast}\,,
\end{align}
where $\widehat{\mathbf{X}}_{np}(\widetilde{w}) = \sum_{S_{np}} \widetilde{w}_j \mathbf{x}_j$ and  $\mathbf{A}_{np}^{\ast}(\widetilde{w}) =
\sum_{S_{np}} \widetilde{w}_j \mathbf{x}_j \mathbf{x}_j^T/\sigma_j^{\ast 2}$. (Note that $\sigma_j^{\ast 2}$ does not have to be the same as
$\widetilde{\sigma}_j^2$ used in constructing the GREG weight in $S_p$.) As in section \ref{sec:estmatch}, $\sigma_j^{\ast 2}$ is often set to
a constant in which case it drops out of the formula for $w_j^{\ast}$. The matched, calibrated estimator is then
\begin{align} \label{eq:YMC}
    \widehat{Y}_{MC} &= \sum_{S_{np}} w_j^{\ast} y_j \notag \\
        &= \sum_{S_{np}} \widetilde{w}_j y_j + \left(\mathbf{X}_U - \widehat{\mathbf{X}}_M \right)^T
        \left[\mathbf{A}_{np}^{\ast}(\widetilde{w})\right]^{-1} \sum_{S_{np}} \widetilde{w}_j \mathbf{x}_j y_j/ \sigma_j^{\ast 2} \notag \\
        &= \widehat{Y}_M + \left( \mathbf{X}_U - \widehat{\mathbf{X}}_M \right)^T \widehat{\mathbf{B}}_{np}^{\ast}(\widetilde{w})\,,
\end{align}
where $\widehat{\mathbf{B}}_{np}^{\ast}(\widetilde{w}) = \left[\mathbf{A}_{np}^{\ast}(\widetilde{w})\right]^{-1} \sum_{S_{np}} \widetilde{w}_j
\mathbf{x}_j y_j/ \sigma_j^{\ast 2}$. As in section \ref{sec:estmatch}, calculations depend on cases (i) and (ii) of the $\widetilde{w}_j$
weights. When case (i) weights are used from $S_p$, the calibrated estimator will be denoted by $\widehat{Y}_{MC1}$; when case (ii) weights are
used, $\widehat{Y}_{MC2}$ denotes the calibrated estimator in subsequent sections.

\subsection{Biases in case (i)} \label{subsec:MCbias.c1}
When $\widetilde{w}_j = \pi_j^{-1}$, $\widehat{\mathbf{X}}_M = \widehat{\mathbf{X}}_{np}(\pi)$ and, after calibration, the estimator of the
total can be written as
\[ \widehat{Y}_{MC1} = \widehat{Y}_{np}(\pi) + \left(\mathbf{X}_U - \widehat{\mathbf{X}}_{np}(\pi) \right)^T
\mathbf{\widehat{B}}_{np}^{\ast}(\pi)\,,\]
where $\widehat{\mathbf{B}}_{np}^{\ast}(\pi)$ is the special case of $\widehat{\mathbf{B}}_{np}^{\ast}(\widetilde{w})$ with $\widetilde{w}_j =
\pi_j^{-1}$. Since $E_{\xi}\left(\widehat{Y}_{np}(\pi) \right) = \widehat{\mathbf{X}}_{np}(\pi) \boldsymbol{\beta}$ under model
\eqref{eq:model} and $E_{\xi}\left( \mathbf{\widehat{B}}_{np}^{\ast}(\pi) \right) = \boldsymbol{\beta}$, $E_{\xi}\left(\widehat{Y}_{MC1} - Y_U
\right)=0$, i.e. $\widehat{Y}_{MC1}$ is $\xi$-unbiased. Thus, calibrating on the $x$'s in the $\xi$-model yields an $\xi$-unbiased estimator
even if $R_j \neq \pi_j$.

To calculate the $R\pi$-expectation, define $\mathbf{B}_U^{\ast} = {\mathbf{A}_U^{\ast}}^{-1} \left(\sum_U \frac{R_j}{\pi_j} \mathbf{x}_j
y_j/\sigma_j^{\ast 2} \right)$ with $\mathbf{A}_U^{\ast} = \sum_U \frac{R_j}{\pi_j} \mathbf{x}_j \mathbf{x}_j^T/\sigma_j^{\ast 2}$.  By the
same type of Taylor series argument as in \citet[sec.6.5]{Sarndal1992},
\begin{equation} \label{eq:YMC1approx}
 \widehat{Y}_{MC1} \doteq \widehat{Y}_{np}(\pi) + \left(\mathbf{X}_U - \widehat{\mathbf{X}}_{np}(\pi) \right)^T \mathbf{B}_U^{\ast}.
\end{equation}
It follows that $E_R E_{\pi}\left(\widehat{Y}_{MC1}\right) = E_R \left(\widehat{Y}_{MC1}\right) \doteq \sum_U R_j y_j / \pi_j +
\left(\mathbf{X}_U - \sum_U R_j \mathbf{x}_j/\pi_j \right)^T \mathbf{B}_U^{\ast}$. If $R_j = \pi_j$, then $\widehat{Y}_{MC1}$ is approximately
$R\pi$-unbiased. Another consequence is that, when $S_{np}$ is calibrated with the $x$'s in model \eqref{eq:model} and $S_p$ has case (i)
weights, $\widehat{Y}_{MC1}$ is $R\pi\xi$-unbiased if $R_j = \pi_j$.

\subsection{Biases in case (ii)} \label{subsec:MCbias.c2}
In case (ii) with $\widetilde{w}_j = g_j/\pi_j$ and $g_j$ defined in \eqref{eq:gj}, the matched estimator after calibration equals
\[ \widehat{Y}_{MC2} = \sum_{S_{np}} g_j^{\ast} g_j y_j/ \pi_j\,,
\]
where
\[ g_j^{\ast} = 1 + \left(\mathbf{X}_U - \widehat{\mathbf{X}}_{np}(\widetilde{w})\right)^T
\left[\widetilde{\mathbf{A}}_{np}^{\ast}(\widetilde{w})\right]^{-1} \mathbf{x}_j/\sigma_j^{\ast 2}\,.
\]
As show in Appendix \ref{app:YMC2approx}, the calibrated estimator of the total is approximately
\begin{equation} \label{eq:YMC2approx1}
    \widehat{Y}_{MC2} \doteq \widehat{Y}_{np}(\pi) \; + \;
        \left(\mathbf{X}_U - \widehat{\mathbf{X}}_p\right)^T \widetilde{\mathbf{B}}_U \; + \;
        \left(\mathbf{X}_U - \widehat{\mathbf{X}}_{np}(\widetilde{w})\right)^T \mathbf{B}_{U}^{\ast}.
\end{equation}

Using \eqref{eq:YMC2approx1}, the $\xi$-expectation is
\[
    E_{\xi}\left(\widehat{Y}_{MC2}\right) \doteq \widehat{\mathbf{X}}_{np}(\pi)\boldsymbol{\beta} +
     \left(\mathbf{X}_U - \widehat{\mathbf{X}}_p\right)^T\boldsymbol{\beta} + \left(\mathbf{X}_U -
     \widehat{\mathbf{X}}_{np}(\widetilde{w})\right)^T \boldsymbol{\beta},
\]
which is not $E_{\xi}(Y_U) = \mathbf{X}_U^T \boldsymbol{\beta}$. That is, $\widehat{Y}_{MC2}$ is $\xi$-biased. This bias occurs even though the
nonprobability sample is calibrated on the $x$'s in the model for $Y$.

If $R_j=\pi_j$, then $E_{R} E_{\pi}\left(\widehat{Y}_{MC2}\right) \doteq Y_U$ and $E_{R} E_{\pi}E_{\xi}\left(\widehat{Y}_{MC2}-Y_U \right)$ is
approximately zero.

The bias results for the matched, calibrated estimators $\widehat{Y}_{MC1}$ and $\widehat{Y}_{MC2}$ can be summarized as follows:
\begin{itemize}
    \item Case (i), $\widetilde{w}_j = \pi_j^{-1}$ and the $\widetilde{w}_j$ are then calibrated to population $x$-totals
    \begin{itemize}
        \item When the linear model \eqref{eq:model} holds, $\widehat{Y}_{MC1}$ is $\xi$-unbiased regardless of whether $R_j = \pi_j$ ;
        \item $\widehat{Y}_{MC1}$ is approximately $R$-, $R\pi$-, and $R\pi\xi$-unbiased in large samples if $R_j = \pi_j$;
    \end{itemize}
    \item Case (ii), $\widetilde{w}_j$ is the GREG weight for its matched unit in $S_p$ and the $\widetilde{w}_j$ are then calibrated to
        population $x$-totals
    \begin{itemize}
       \item $\widehat{Y}_{MC2}$ is $\xi$-biased even if \eqref{eq:model} holds and the nonprobability sample $S_{np}$ is calibrated on the
           $x$'s in the model;
       \item $\widehat{Y}_{MC2}$ is approximately $R\pi$-unbiased in large samples if $R_j = \pi_j$;
       \item $\widehat{Y}_{MC2}$ is approximately $R\pi\xi$-unbiased in large samples when \eqref{eq:model} holds if $R_j = \pi_j$;
    \end{itemize}
\end{itemize}
If case (i) holds where the weights assigned to matched units are inverses of selection probabilities from $S_p$, the situation is more
straightforward than case (ii). $R$-unbiasedness in case (i) requires that the pseudo-inclusion probabilities can be taken from the probability
sample, i.e., $R_j = \pi_j$. Nonetheless, in case (i) calibrating the nonprobability sample does produce an $\xi$-unbiased estimator even if
$R_j \neq \pi_j$, as one would hope. However, in case (ii) when the weights from the probability sample are calibrated and the nonprobability
sample is further calibrated on the same $x$'s, the resulting estimator is not $\xi$-unbiased.

\subsection{Variance of the Matched, Calibrated Estimator in case (i)} \label{subsec:vmc1}
To compute the $\xi$-model variance, note that the estimator of total can also be written as $\widehat{Y}_{MC1} = \sum_{S_{np}}
g_j^{\ast}y_j/\pi_j$ with $g_j^{\ast}$ defined in \eqref{eq:wjstar} with $\widetilde{w}_j = 1/\pi_j$. The $\xi$-variance is then
\begin{equation*}
    V_{\xi}\left(\widehat{Y}_{MC1} \right) = \sum_{S_{np}} \left(\frac{g_j^{\ast}}{\pi_j} \right)^2 \sigma_j^2\,.
\end{equation*}
It follows that the $R\pi\xi$-variance is $V_{R\pi\xi}\left(\widehat{Y}_{MC1} \right) = \sum_{U} \left(g_j^{\ast 2}/ \pi_j \right) \sigma_j^2$.
The $\xi$-variance can be estimated with
\begin{equation}\label{eq:Vxi.YMChat.c1}
    v_{\xi}\left(\widehat{Y}_{MC1}\right) = \sum_{S_{np}} \left(\frac{g_j^{\ast}}{\pi_j} \right)^2 \widehat{e}_j^{\ast 2}\,,
\end{equation}
where $\widehat{e}_j^{\ast 2} = y_j - \mathbf{x}_j^T \widehat{\mathbf{B}}_{np}^{\ast}(\pi)$ with $\widehat{\mathbf{B}}_{np}^{\ast}(\pi) =
\left(\sum_{S_{np}} \mathbf{x}_j \mathbf{x}_j^T / (\pi_j \sigma_j^{\ast 2})   \right)^{-1} \sum_{S_{np}} \mathbf{x}_j y_j / (\pi_j
\sigma_j^{\ast 2})$.

To compute the $R$- and $R\pi$-variance, we use the approximation in \eqref{eq:YMC1approx}. Assume that $R_j = \pi_j$ so that
$\widehat{Y}_{MC1}$ is $R$-unbiased. Based on results in section \ref{subsec:MCbias.c1}, the estimator can be approximated as
\begin{align} \label{eq:YMCapprox.c1}
 \widehat{Y}_{MC1} & = \widehat{Y}_{np}(\pi) + \left(\mathbf{X}_U - \widehat{\mathbf{X}}_{np}(\pi) \right)^T
 \widehat{\mathbf{B}}_{np}^{\ast}(\pi) \notag \\
    & \doteq \widehat{Y}_{np}(\pi) + \left(\mathbf{X}_U - \widehat{\mathbf{X}}_{np}(\pi) \right)^T \mathbf{B}_U^{\ast} \notag \\
    & = \sum_{S_{np}} \pi_j^{-1} e_j^{\ast} + \mathbf{X}_U^T \mathbf{B}_U^{\ast}\,,
\end{align}
where $e_j^{\ast} = y_j - \mathbf{x}_j^T \mathbf{B}_U^{\ast}$. The $R$- (and $R\pi$-) variance is, thus, equal to the variance of the first
term in the last line of \eqref{eq:YMCapprox.c1}. If the sample $S_{np}$ is treated as being selected with replacement, then a variance
estimator is
\begin{align} \label{eq:VRpi.YMC1}
    v_{R\pi}\left(\widehat{Y}_{MC1} \right)
        &= \frac{n}{n-1} \sum_{j \in S_{np}} \left(\widetilde{w}_{j}\widehat{e}_j^{\ast} - \frac{1}{n} \sum_{j^{\prime} \in S_{np}}
        \widetilde{w}_{j^\prime}\widehat{e}_{j\prime}{\ast}\right)^2.
\end{align}

\subsection{Variance of the Matched, Calibrated Estimator in case (ii)} \label{subsec:vmc2}
As shown in Appendix \ref{app:YMC2var}, approximation \eqref{eq:YMC2approx1} can be rewritten as
\[
    \widehat{Y}_{MC2} \doteq \sum_{S_{np}} y_j \left( \frac{1}{\pi_j} +  F_j \right)
    + \sum_{U-S_{np}} y_j F_j\,,
\]
where $F_j$ is a term that is $O_p\left(n^{-1/2} \right)$. As a result, $V_{\xi}\left( \widehat{Y}_{MC2}\right) \doteq \sum_{S_{np}}
\left(\sigma_j / \pi_j \right)^2$, which can be estimated with
\begin{equation} \label{eq:vxi.MC2c2}
    v_{\xi}\left(\widehat{Y}_{MC2}\right) = \sum_{S_{np}} \left( \frac{\widehat{e}_j^{\ast}}{\pi_j} \right)^2.
\end{equation}

Rewriting \eqref{eq:YMC2approx1}, the calibrated estimator of the total is also
\begin{align} \label{eq:YMCapprox.v2}
    \widehat{Y}_{MC2} & \doteq \sum_{S_{np}} \frac{e_j^{\ast}}{\pi_j} + \left(\mathbf{X}_U - \widehat{\mathbf{X}}_p \right)^T
    \widetilde{\mathbf{B}}_U + \mathbf{X}_U^T \mathbf{B}_U^{\ast}\,,
\end{align}
where $e_j^{\ast}$ was defined above. Using the total variance formula, the $R\pi$-variance can be derived as
\begin{align*}
    V_{R\pi}\left(\widehat{Y}_{MC2}\right) &= V_R E_{\pi}\left(\widehat{Y}_{MC2} \mid S_{np} \right) + E_R V_{\pi}\left(\widehat{Y}_{MC2} \mid
    S_{np}\right) \\
    &=
    V_R\left(\sum_{S_{np}} \frac{e_j^{\ast}}{\pi_j} \right) + E_R V_{\pi}\left[\left( \mathbf{X}_U - \widehat{\mathbf{X}}_p \right)^T
    \widetilde{\mathbf{B}}_U \right] \\
    &= V_R\left(\sum_{S_{np}} \frac{e_j^{\ast}}{\pi_j} \right) +
     \widetilde{\mathbf{B}}_U^T V_{\pi}\left( \widehat{\mathbf{X}}_p \right) \widetilde{\mathbf{B}}_U\,.
\end{align*}
An estimator of this variance is
\begin{equation}\label{eq:vR.YMC2}
    v_{R\pi}\left(\widehat{Y}_{MC2} \right) =  v_R\left(\sum_{S_{np}} \frac{e_j^{\ast}}{\pi_j} \right) +
    \widehat{\widetilde{\mathbf{B}}}_{np}(\pi)^T v_{\pi}\left(\widehat{\mathbf{X}}_p \right) \widehat{\widetilde{\mathbf{B}}}_{np}(\pi)
\end{equation}
with $v_{R}\left(\sum_{S_{np}} \frac{e_j^{\ast}}{\pi_j} \right)$ being a variance estimator of an estimated total appropriate to how the
nonprobability sample is handled. We use $\widehat{\widetilde{\mathbf{B}}}_{np}(\pi)$ in \eqref{eq:vR.YMC2} rather than an estimator with
$\widetilde{w}$ weights since the former is expected to be somewhat more stable. If $S_{np}$ is treated as being with-replacement, the first
component in \eqref{eq:vR.YMC2} can be computed with
\eqref{eq:VRpi.YMC1}.

Details of calculating $V_{R\pi\xi}\left(\widehat{Y}_{MC2} \right)$ are in Appendix \ref{app:YMC2var}.  This variance can be estimated with
\begin{equation} \label{eq:vRpixi.YMC2}
    v_{R\pi\xi}\left(\widehat{Y}_{MC2} \right) = \sum_{S_{np}} \left(\frac{\widehat{e}_j^{\ast}}{\pi_j}\right)^2 +
    \widehat{\widetilde{\mathbf{B}}}_{np}(\pi)^T v_{\pi}\left({\widehat{\mathbf{X}}_p} \right) \widehat{\widetilde{\mathbf{B}}}_{np}(\pi)\,.
\end{equation}

For each of the variance estimators above for the matched, calibrated estimator in cases (i) and (ii), it is important to remember that unless
$R_j = \pi_j$ the estimator of total itself will be biased. If so, the mean square error will have a bias-squared component that none of the
variance estimators will reflect.

In the combination above, both the weights in $S_p$ and those in $S_{np}$ are calibrated to a given set of $x$'s. This is similar to the
situation studied by \citet[p.368]{Rao.2002}, who noted that in a regression with calibration weights, GREG residuals are based on the
regression of model residuals on $\mathbf{X}$. If the model fits well, there will be very little association between those residuals and
$\mathbf{X}$ leading to no gain compared to an estimator not using calibration weights. In our situation, when the estimators of totals are
unbiased, we can expect $\widehat{Y}_{M2}$ with calibration in $S_{p}$, $\widehat{Y}_{MC1}$ with no calibration in $S_{p}$ and calibration in
$S_{np}$, and $\widehat{Y}_{MC2}$ with calibration in both $S_p$ and $S_{np}$ to be about equally precise---a point borne out by the simulation
in section \ref{sec:sim}.

\section{Simulation Studies} \label{sec:sim}
To study the performance of the proposed estimators described above,
we performed two simulation studies with an artificial population. In the first, conditions are created where close matches can be found
between units in the probability sample and the nonprobability sample. In the second simulation, close matches are much less likely.

\subsection{Simulation Study I}

In the simulation, a finite population of size $N=100,000$ was based on the following model:
\begin{equation*}
  E_{\xi}(Y) = \alpha + \beta X, V_{\xi}(Y) = \sigma^2 X^{3/2}\,,
\end{equation*}
where $\alpha=0.4$, $\beta=0.25$, $\sigma^2=0.0625$,
and $X$ follows a gamma distribution
with density function $f(x) = 0.04 x \exp( - x/5)$. This is the same model as used by \citet{Hansen.1983}; the function \texttt{HMT} in the R
\texttt{PracTools} package \citep{VDK.2020} was used to generate the population.
Conditional on $X$, $Y$ follows a gamma distribution with
density function  $g(y; x) = (1/b^{c}\Gamma(c)) y^{c-1} \exp( -y/b)$,
where $b = 1.25 x^{3/2} (8 + 5x)^{-1} $, $c =0.04 x^{-3/2} (8 + 5x)^{2}$ and $\Gamma(\cdot)$ is the gamma function.
The finite population is stratified into five strata by ranges of the variable $X$,
such that each stratum has approximately the same total of $X$.
A stratified, probability sample $S_{p}$ of size $n=250$ is then selected from the population
using stratified, simple random sampling (\textit{stsrs}) without replacement,
in which the sample stratum sizes are given by $(50, 50, 50, 50, 50)$.
Further, a stratified, volunteer panel of size $M = 1250$
is selected from the population with stratum sample sizes $(250, 250, 250, 250, 250)$
using stratified, simple random sampling. Although the volunteer panel is a probability sample,
their weights are treated as unknown for the simulation. Note that the sampling fractions of both $S_p$ and $S_{np}$ are small and, thus, will
not affect the empirical variances of estimates.

For each unit of the probability sample of $n=250$, we find the closest matching unit of the volunteer panel to obtain the matched,
non-probability sample $S_{np}$ of size $n=250$,
using single nearest neighbor matching without replacement based on the single auxiliary variable $X$.
The units in the volunteer panel are then assigned the weight of their nearest neighbor match from the probability sample using the R package
\texttt{Matching} \citep{Sekhon.2011}.
In this example, finding close matches is fairly easy, and we should have $R_{j}=\pi_{j}, j \in S_{np}$, in almost all cases because both $S_p$
and $S_{np}$ are \textit{stsrs}.
The parameter of interest is the population total of $Y$.
Finally, the matched estimator and the matched, calibrated estimator under cases $(i)$ and $(ii)$ in section \ref{sec:estmatch} are computed,
denoted by
\begin{itemize}
    \item $\widehat{Y}_{M1}$, estimator \eqref{eq:est.tot} with $1/\pi$ weights from the matched units in $S_p$,
    \item $\widehat{Y}_{M2}$, estimator \eqref{eq:YM.greg} with GREG weights from the matched units in $S_p$,
    \item $\widehat{Y}_{MC1}$, estimator \eqref{eq:YMC} with $1/\pi$ weights from the matched units in $S_p$ followed by calibration in
        $S_{np}$, and
    \item $\widehat{Y}_{MC2}$, estimator \eqref{eq:YMC} with GREG weights from the matched units in $S_p$ followed by  calibration in
        $S_{np}$.
\end{itemize}
The above process is repeated 5000 times.
The percentage relative biases (relbiases), the variances and the mean squared errors of the matched estimator and the matched,
calibrated estimator under cases $(i)$ and $(ii)$, are presented in Table 1.
The empirical percent relative bias is defined as $ 100 \times bias(\widehat{Y})/Y$.

For comparison we included a doubly robust estimator, denoted by $\hat{Y}_{DR}$, that was computed without matching. This estimator was
computed in two steps as described in \citet{Elliott.2017}. First, an equal probability subsample of $n = 250$ was selected from the volunteer
panel of $m = 1250$.  Then, $S_p$ and $S_{np}$ are combined. Units in $S_{np}$ are given a weight of 1 while units in $S_p$ were assigned their
sampling weight of $1/\pi_i$. A logistic regression with $X$ as the covariate was run to predict the probability of being in $S_{np}$. The
weight for unit $j$ in $S_{np}$ was then calculated as $w_j = (1 - \hat{R}_j)/\hat{R}_j$ where $\hat{R}_j$ is the predicted probability of
being in $S_{np}$ \citep[see][]{Wang.2021}. Without the odds transformation, the estimator would be somewhat biased \citep{CLW.2020}, but in
this case the bias was negligible since $S_{np}$ is a small fraction of the population \citep{Wang.2021}. Finally, the estimator was calibrated
with a model having an intercept and $X$.

\singlespacing
{\centering
Table 1~~{Simulation Study I: Monte Carlo percent relative biases, variances and mean squared errors of the point estimators}}
\begin{longtable}{p{.12\linewidth}p{.18\linewidth}p{.18\linewidth}p{.18\linewidth}p{.22\linewidth}}
\toprule
 \hfil Estimators \hfil  & \hfil Relative Bias \hfil & \hfil Variance \hfil & \hfil MSE \hfil & \hfil Ratio to \hfil \\
 \hfil & \hfil (\%) \hfil &\hfil ($\div 10^7$) \hfil & \hfil ($\div 10^7$) \hfil & \hfil min MSE \hfil \\
\midrule
\hfil $\widehat{Y}_{M1}$   \hfil & \hfil    -0.0044	&	\hfil	9.04	&	\hfil	9.04	&	\hfil	 1.16	 \\
\hfil $\widehat{Y}_{M2}$   \hfil & \hfil    0.0241	&	\hfil	7.79	&	\hfil	7.79	&	\hfil	 1.00	 \\
 \hfil $\widehat{Y}_{MC1}$  \hfil & \hfil   0.0029	&	\hfil	7.79	&	\hfil	7.79	&	\hfil	 1.00	 \\
 \hfil $\widehat{Y}_{MC2}$   \hfil & \hfil  0.0030	&	\hfil	7.79	&	\hfil	7.79	&	\hfil	 1.00	 \\
 \hfil $\widehat{Y}_{DR}$   \hfil & \hfil   -0.0671	&	\hfil	9.39	&	\hfil	9.39	&	\hfil	 1.21	 \\
\bottomrule
\end{longtable}
\doublespacing

Simulation results in Table 1 show that the absolute relative biases of the matched estimators
under the two cases of weights from $S_p$ are small and close to those of the corresponding
matched, calibrated estimators under the two cases.
Thus, both the matched estimators and the matched, calibrated estimators are unbiased
when $R_{j} \cong \pi_{j}, j \in S_{np}$ in both cases $(i)$ and $(ii)$ as predicted by the theory in sections \ref{subsec:bias.c1} and
\ref{subsec:bias.c2}. The variances and MSEs of $\hat{Y}_{M2}$, $\hat{Y}_{MC1}$, and $\hat{Y}_{MC2}$ are all equal as anticipated in the
comment at the end of section \ref{subsec:vmc2} and are about 16\% smaller than those of $\hat{Y}_{M1}$. Consequently, while all estimates are
approximately unbiased, the calibration adjustment after matching produces more efficient estimators compared to only matching under case
$(i)$. Also noteworthy is the fact that the doubly robust estimator, $\hat{Y}_{DR}$, has a 21\% larger MSE than the best of the matching
estimates. This is a consequence of the logistic model used to estimate $Pr\left(j \in S_{np} \right)$ being a misspecification.

In addition to the point estimators, the variance estimators of the matched estimator and the matched,
calibrated estimator under cases $(i)$ and $(ii)$ are also computed
according to equations \eqref{eq:VR.hat}, \eqref{eq:VxM1c1}, \eqref{eq:vRpixi.M1c1}, \eqref{eq:vxi.M2c2}, \eqref{eq:vRpi.M2c2},
\eqref{eq:vRpixi.M2c2}, \eqref{eq:Vxi.YMChat.c1}, \eqref{eq:VRpi.YMC1}, \eqref{eq:vxi.MC2c2}, \eqref{eq:vR.YMC2}, and \eqref{eq:vRpixi.YMC2}.
In all cases $S_{np}$ is treated as an unstratified, with replacement sample. Percent relative biases (RB) are computed for the variance
estimators with respect to the empirical variances (Empvar) and MSEs of the point estimators
across the 5000 simulations:
\begin{equation*}
   RB.Empvar = \frac{100 \times {\big(\sum^{B}_{b=1} v^{(b)}(\widehat{Y}) / B - V(\widehat{Y})\big)}}{ V(\widehat{Y})}\,,
\end{equation*}
\begin{equation*}
    RB.MSE = \frac{100 \times {\big(\sum^{B}_{b=1} v^{(b)}(\widehat{Y}) / B - MSE(\widehat{Y})\big)}}{ MSE(\widehat{Y})}\,,
\end{equation*}
where $V(\widehat{Y})$ is the empirical or monte carlo variance of a point estimator $\widehat{Y}$,
$MSE(\widehat{Y})$ is MSE of the point estimator $\widehat{Y}$,
$v^{(b)}(\widehat{Y})$ is a variance estimator of $\widehat{Y}$ computed from the $b^{th}$ simulated sample, and
$B=5000$ is the total number of simulation runs.
The percent relative biases (RB) and 95\% confidence interval (CI) coverages
using the normal approximation and the different variance estimates,
are presented in Table 2.

With three exceptions, the relbiases in Table 2 are small, ranging from -2.2\% to 3.1\%. An exception is $v_{\xi}(\widehat{Y}_{M1})$ which is a
15.8\% underestimate due to the fact that it does not account for the variability of $\hat{X}_{np}$ as shown in section \ref{subsec:Vcase.i}.
The $R\pi$ and $R\pi\xi$ estimators for $\widehat{Y}_{M2}$ are about 22\% overestimates. As explained in Appendix \ref{app:YM2c2var}, these
estimators will not fully account for precision gains due to calibration of weights in $S_p$ when the $x$-matches are extremely close.
Confidence interval coverage ranges from 94.6\% to 96.7\% except for $v_{\xi}(\widehat{Y}_{M1})$ which covers in 92.3\% of samples due to its
underestimation.

\singlespacing
{\centering
Table 2~~{Simulation Study I: Percent relative biases and 95\% confidence interval coverages of the variance estimators}}
\begin{longtable}{p{.14\linewidth}p{.18\linewidth}p{.18\linewidth}p{.18\linewidth}p{.18\linewidth}}
\toprule
\hfil Estimators \hfil &  \hfil RB.Empvar (\%) \hfil & \hfil RB.MSE (\%) \hfil & \hfil CI coverage (\%)  \hfil \\
\midrule
\hfil $v_{\xi}(\widehat{Y}_{M1})$     \hfil & \hfil     -15.8	\hfil & \hfil	-15.8	\hfil & \hfil	 92.3	 \hfil \\
\hfil $v_{R}(\widehat{Y}_{M1})$       \hfil & \hfil       3.1	\hfil & \hfil	3.1	    \hfil & \hfil	 94.6	 \hfil \\
\hfil $v_{R\pi\xi}(\widehat{Y}_{M1})$ \hfil & \hfil  2.6	\hfil & \hfil	2.7	    \hfil & \hfil	 96.4	 \hfil	 \\
\hfil $v_{\xi}(\widehat{Y}_{M2})$     \hfil & \hfil      -2.1	\hfil & \hfil	-2.1	\hfil & \hfil	 94.6	 \hfil \\
\hfil $v_{R\pi}(\widehat{Y}_{M2})$    \hfil & \hfil    22.5	\hfil & \hfil	22.6	\hfil & \hfil	 96.5	 \hfil	 \\
\hfil $v_{R\pi\xi}(\widehat{Y}_{M2})$ \hfil & \hfil 22.0	\hfil & \hfil	22.0	\hfil & \hfil	 96.7	 \hfil	 \\
\hfil $v_{\xi}(\widehat{Y}_{MC1})$    \hfil & \hfil    -2.1	\hfil & \hfil	-2.1	\hfil & \hfil	 94.7	 \hfil	 \\
\hfil $v_{R}(\widehat{Y}_{MC1})$      \hfil & \hfil      -1.8	\hfil & \hfil	-1.8	\hfil & \hfil	 94.6	 \hfil \\
\hfil $v_{\xi}(\widehat{Y}_{MC2})$    \hfil & \hfil    -2.2	\hfil & \hfil	-2.2	\hfil & \hfil	 94.6	 \hfil	 \\
\hfil $v_{R\pi}(\widehat{Y}_{MC2})$    \hfil & \hfil    1.0	\hfil & \hfil	1.1	    \hfil & \hfil	 94.9	 \hfil	 \\
\hfil $v_{R\pi\xi}(\widehat{Y}_{MC2})$ \hfil & \hfil     0.6	\hfil & \hfil	0.7	    \hfil & \hfil	 94.9	 \hfil	 \\
\bottomrule
\end{longtable}
\doublespacing

\subsection{Simulation Study II}

In this simulation, we consider a case in which $R_{j} \neq \pi_{j}, j \in S_{np}$.
The same finite population of size $N=100,000$ is used as in simulation study I along with
a stratified, probability sample $S_{p}$ of size $n=250$.
A volunteer panel of expected size $m = 1250$ is selected from the population
using Poisson sampling with selection probabilities $\pi'_{i}$ defined as follows:
\begin{equation*}
   \pi_{i} = 0.085 \exp(-0.085 X)\,,
\end{equation*}
\begin{equation*}
     \pi'_{i}=\frac{ m \pi_{i}}{\sum^{N}_{i=1}\pi_{i}}\,.
\end{equation*}
With this definition of $\pi_i$, the probability of being in $S_{np}$ decreases with increasing $X$. This kind of selection for the volunteer
panel will generally result in $R_{j} \neq \pi_{i}$, for a unit $j \in S_{np}$ matched to a unit $i \in S_p$.

As in simulation I, single nearest neighbor matching without replacement based on the variable $X$ is adopted to conduct matching for the
probability sample.
The matched estimator, the matched, calibrated estimator and their variance estimators under cases $(i)$ and $(ii)$ are computed.
The above procedure is repeated 5000 times.
The relative biases, the variances and the mean squared errors are listed in Table 3. Also, the same relative biases and 95\% CI coverages of
variance estimators as those in simulation study I are displayed in Table 4.

In Table 3 the matched estimators, $\widehat{Y}_{M1}$ and $\widehat{Y}_{M2}$, have biases of about -5\%. These biases are largely corrected by
calibrating with $\widehat{Y}_{MC1}$ and $\widehat{Y}_{MC2}$.  The calibrated estimates, consequently, have substantially smaller MSEs than
$\widehat{Y}_{M1}$ and $\widehat{Y}_{M2}$ because of their reduced bias. The doubly robust estimator, $\widehat{Y}_{DR}$, is also approximately
unbiased; however, its variance and MSE are 50\% higher than those of $\widehat{Y}_{MC1}$ and $\widehat{Y}_{MC2}$.

\singlespacing
{\centering
Table 3~~{Simulation Study II: Percent relative biases, variances and mean squared errors of the point estimators}}
\begin{longtable}{p{.12\linewidth}p{.18\linewidth}p{.18\linewidth}p{.18\linewidth}p{.22\linewidth}}
\toprule
 \hfil Estimators \hfil  & \hfil Relative Bias \hfil & \hfil Variance \hfil & \hfil MSE \hfil & \hfil Ratio to \hfil \\
 \hfil & \hfil (\%) \hfil &\hfil ($\div 10^7$) \hfil & \hfil ($\div 10^7$) \hfil & \hfil min MSE \hfil \\
\midrule
\hfil $\widehat{Y}_{M1}$   \hfil & \hfil     -5.2   \hfil & \hfil    8.86  \hfil & \hfil 31.11 \hfil & \hfil  3.9 \\
\hfil $\widehat{Y}_{M2}$   \hfil & \hfil     -5.1   \hfil & \hfil    7.96  \hfil & \hfil 29.84 \hfil & \hfil  3.8 \\
 \hfil $\widehat{Y}_{MC1}$  \hfil & \hfil    -0.2   \hfil  & \hfil   7.89  \hfil & \hfil  7.91 \hfil & \hfil  1.0 \\
 \hfil $\widehat{Y}_{MC2}$   \hfil & \hfil   -0.2   \hfil  & \hfil   7.89  \hfil & \hfil  7.91 \hfil & \hfil  1.0 \\
 \hfil $\widehat{Y}_{DR}$   \hfil & \hfil    -0.2   \hfil  & \hfil   11.93 \hfil & \hfil 11.95 \hfil & \hfil  1.5 \\
\bottomrule
\end{longtable}
\doublespacing

In Table 4 the variance estimates for $\widehat{Y}_{M1}$ and $\widehat{Y}_{M2}$ are biased estimates of the empirical variance and severe
underestimates of the MSEs. This leads to CIs that cover only about 56\% to 67\% of the time for the first four variance estimates in Table 4.
Since $v_{R\pi}(\widehat{Y}_{M2})$ and $v_{R\pi\xi}(\widehat{Y}_{M2})$ overestimate the empirical variances by about 23\%, their CIs do cover
the population totals in 96.7\% of samples. The fact that calibrating removes the bias of the matching estimators plus the low biases of the
variance estimators for $\widehat{Y}_{MC1}$ and $\widehat{Y}_{MC2}$ leads to CI coverage of 93.6\% to 94.4\%.

\singlespacing
{\centering
Table 4~~{Simulation Study II: Percent relative biases and 95\% confidence interval coverages of the variance estimators}}
\begin{longtable}{p{.15\linewidth}p{.18\linewidth}p{.18\linewidth}p{.18\linewidth}p{.18\linewidth}}
\toprule
\hfil Estimators \hfil &  \hfil RB.Empvar (\%) \hfil & \hfil RB.MSE (\%) \hfil & \hfil CI coverage (\%)  \hfil \\
\midrule
\hfil $v_{\xi}(\widehat{Y}_{M1})$      \hfil & \hfil   -18.7     \hfil & \hfil -76.8  \hfil & \hfil  56.4   \hfil  \\
\hfil $v_{R}(\widehat{Y}_{M1})$        \hfil & \hfil   8.1     \hfil & \hfil   -69.2  \hfil & \hfil  65.2   \hfil  \\
\hfil $v_{R\pi\xi}(\widehat{Y}_{M1})$  \hfil & \hfil   7.7     \hfil  & \hfil  -69.3  \hfil & \hfil  67.7   \hfil \\
\hfil $v_{\xi}(\widehat{Y}_{M2})$      \hfil & \hfil   -9.3   \hfil  & \hfil   -75.8  \hfil & \hfil  56.9   \hfil \\
\hfil $v_{R\pi}(\widehat{Y}_{M2})$      \hfil & \hfil  23.2  \hfil & \hfil     -67.2  \hfil & \hfil  96.7   \hfil  \\
\hfil $v_{R\pi\xi}(\widehat{Y}_{M2})$   \hfil & \hfil  22.8   \hfil & \hfil    -67.3  \hfil & \hfil  96.7   \hfil  \\
\hfil $v_{\xi}(\widehat{Y}_{MC1})$      \hfil & \hfil  -1.7    \hfil & \hfil   -1.9   \hfil & \hfil  94.4   \hfil  \\
\hfil $v_{R}(\widehat{Y}_{MC1})$         \hfil & \hfil -8.3  \hfil  & \hfil    -8.5   \hfil & \hfil  93.7   \hfil \\
\hfil $v_{\xi}(\widehat{Y}_{MC2})$     \hfil & \hfil   -8.7  \hfil  & \hfil    -8.9   \hfil & \hfil  93.6   \hfil \\
\hfil $v_{R\pi}(\widehat{Y}_{MC2})$     \hfil & \hfil   -5.5  \hfil  & \hfil    -5.8   \hfil & \hfil  94.0   \hfil \\
\hfil $v_{R\pi\xi}(\widehat{Y}_{MC2})$  \hfil & \hfil   -5.9  \hfil  & \hfil    -6.1   \hfil & \hfil  94.0   \hfil \\
\bottomrule
\end{longtable}
\doublespacing

\section{Illustration with Real Population} \label{sec:realdata}

To further assess the performance of the matching estimators, they are applied to data obtained from the 2015 US Behavioral Risk Factor
Surveillance Survey (http://www.cdc.gov/BRFSS), which is a sample from the US population 18 years and older.  The file contains information
about whether persons used the internet in the past 30 days (\texttt{INTERNET}). The BRFSS is part of a national state-by-state system of
surveys used to monitor health conditions in the United States. Data are collected through telephone household interviews. The analytic
variables $Y$ in this study are whether respondents were ever diagnosed with a heart attack (\texttt{CVDINFR4}),
were ever told by a medical
professional that they have diabetes (\texttt{DIABETE3}), and were ever told they had a stroke (CVDSTRK3). Although each of these analysis
variables is binary, use of linear estimators, as studied in previous sections, is standard survey practice, largely because of their
convenience for data analysts.

Covariates associated with $Y$ are sex, age, race, marital status, physical weight, employment status, education level, income level, whether
respondents smoked at least 100 cigarettes in their entire life, and whether respondents participated in any physical activities or exercises
in the past 30 days in 2015. All of the variables are shown in Table 5.

After deleting cases with either a missing, a don't know or a refused response to any of these variables, 315,669 persons are available for
this study. Two weights are provided with the dataset: \texttt{X\_WT2RAKE}, which is a design weight and \texttt{X\_LLCPWT}, which is a raked,
final weight. According to the documentation (\url{https://www.cdc.gov/brfss/annual_data/2017/pdf/weighting-2017-508.pdf}), BRFSS rakes the
design weight to eight margins (gender by age group, race/ethnicity, education, marital status, tenure, gender by race/ethnicity, age group by
race/ethnicity, and phone ownership). The raking also serves as a noncoverage/nonresponse adjustment. Because of the asymptotic equivalence of
the GREG and raked estimators shown by \citet{Deville.1992}, the earlier theory in sections \ref{sec:estmatch} and \ref{sec:calib} should apply
to estimators based on \texttt{X\_LLCPWT}.

\singlespacing
{\centering
Table 5~~{Covariates used in the BRFSS simulation study}}\vspace{1ex}
\begin{longtable}{p{.13\linewidth}p{.13\linewidth}p{.65\linewidth}}
\toprule
\hfil Variables \hfil & \hfil Type \hfil & \hfil Description \hfil \\
\midrule
 SEX      & 2 categories  &  Respondents sex: 1=Male; 2=Female  \\
 X\_AGE  & 6 categories  &  Imputed age in six groups: 1=Age 18 to 24; 2=Age 25 to 34; 3=Age 35 to 44; 4=Age 45 to 54; 5=Age 55 to 64; 6=Age 65
 or older  \\
 X\_RACE   & 8 categories  &   Computed race-ethnicity grouping: 1=White only, non-Hispanic; 2=Black only, non-Hispanic; 3=American Indian or
 Alaskan Native only, Non-Hispanic; 4=Asian only, non-Hispanic; 5=Native Hawaiian or other Pacific Islander only, Non-Hispanic; 6=Other race
 only, non-Hispanic; 7=Multiracial, non-Hispanic; 8=Hispanic  \\
 MARITAL  & 6 categories  &  Marital status: 1=Married; 2=Divorced; 3=Widowed; 4=Separated; 5=Never married; 6=A member of an unmarried couple
 \\
 WEIGHT2  & Continuous    &  Reported weight in pounds: 50-999    \\
 EMPLOY1  & 8 categories  &  Employment status: 1=Employed for wages; 2=Self-employed; 3=Out of work for 1 year or more; 4=Out of work for less
 than 1 year; 5=A homemaker; 6=A student; 7=Retired; 8=Unable to work \\
 EDUCA    & 6 categories  &  Education level: 1=Never attended school or only kindergarten; 2=Grades 1 through 8 (Elementary); 3= 	 Grades 9
 through 11 (Some high school); 4=Grade 12 or GED (High school graduate); 5=College 1
 year to 3 years (Some college or technical school); 6=College 4 years or more (College graduate)   \\
 INCOME2  & 8 categories  &  Income level: 1=Less than \$10,000; 2=\$10,000 to less than \$15,000; 3=\$15,000 to less than \$20,000; 4=\$20,000
 to less than \$25,000; 5=\$25,000 to less than \$35,000; 6=\$35,000 to less than
 \$50,000; 7=\$50,000 to less than \$75,000; 8=\$75,000 or more  \\
 SMOKE100 & 2 categories  &  Smoked at least 100 cigarettes?: 1=Yes; 2=No \\
 EXERANY2 & 2 categories  &  Exercise in past 30 days?: 1=Yes; 2=No   \\
 INTERNET & 2 categories  &  Internet use in the past 30 days?: 1=Yes; 2=No  \\
 CVDINFR4 & 2 categories  &  Ever diagnosed with heart attack?: 1=Yes; 2=No  \\
 DIABETE3 & 2 categories  &  Ever told you have diabetes?: 1=Yes; 2=No  \\
 CVDSTRK3 & 2 categories  &  Ever told you had a stroke?: 1=Yes; 2=No  \\
\bottomrule
\end{longtable}
\doublespacing

In this dataset of 315,669 persons, 256,949 people who had used the internet in the past 30 days are considered as the web (nonprobability)
subset. Using the \texttt{X\_LLCPWT} weights, the web population is only 84\% (81\% unweighted) of the target population, indicating that the
effect of coverage error could be substantial. Moreover, the weighted distributions of the categorical covariates among all respondents in the
web, non-web, and full populations are given in Table 6. Categories of some variables are combined in Table 6 and in the simulation compared to
the categories in Table 5 because they are small. Table 7 gives the proportions that reported a heart attack, diabetes, or a stroke in the web,
non-web, and full populations.


\singlespacing
\small
{\centering
Table 6~~{Distributions of the categorical variables and means of the continuous variable, body weight, in the web, non-web, and full
populations}}\vspace{1ex}
\begin{longtable}{p{.13\linewidth}p{.35\linewidth}p{.15\linewidth}p{.15\linewidth}p{.15\linewidth}}
\toprule
\multicolumn{2}{c}{Variables} & Web Population & Non-web Population & Target Population\\
\midrule
 SEX        & Male   & 0.50	&	0.49	&	0.50\\
            & Female & 0.50	&	0.51	&	0.50 \\
            & \\
 X\_AGE  & Age 18 to 24     &	0.13	&	0.02	&	0.11 \\
            & Age 25 to 34     &	0.20	&	0.05	&	0.18 \\
            & Age 35 to 44     &	0.19	&	0.09	&	0.17 \\
            & Age 45 to 54     &	0.19	&	0.16	&	0.18 \\
            & Age 55 to 64     &	0.16	&	0.22	&	0.17 \\
            & Age 65 or older  &	0.14	&	0.45	&	0.19 \\
            & \\
 X\_RACE   & Non-black, non-Hispanic                 & 0.90	&	0.84	&	0.89 \\
           & Black only, non-Hispanic                & 0.10	&	0.16	&	0.11 \\
             & \\
 MARITAL  & Married or member of an unmarried couple & 0.59	&	0.47	&	0.58  \\
          & Divorced                                 & 0.11	&	0.15	&	0.11  \\
          & Widowed, separated, never married        & 0.30	&	0.38	&	0.31  \\
          & \\
 EMPLOY1  & Employed for wages, self-employed        &	0.65	&	0.30	&	0.59  \\
          & Out of work                              &	0.05	&	0.06	&	0.05  \\
          & Other (homemaker, student, retired, unable to work)  &	0.30	&	0.63	&	0.35  \\
          & \\
 EDUCA    & Grade 11 or less            & 0.08	&	0.39	&	0.13   \\
          & Grade 12 or equivalent      & 0.25	&	0.37	&	0.27   \\
          & College 1 year to 3 years   & 0.35	&	0.18	&	0.32   \\
          & College 4 years or more     & 0.33	&	0.06	&	0.29   \\
          & \\
 INCOME2  &  Less than \$25,000              &	0.21	&	0.60	&	0.27   \\
          & \$25,000 to less than \$50,000   &	0.24	&	0.27	&	0.24   \\
          & \$50,000 to less than \$75,000   &	0.17	&	0.07	&	0.16  \\
          & \$75,000 or more                 &	0.38	&	0.06	&	0.33  \\
          & \\
 SMOKE100 & Smoked at least 100 cigarettes       & 0.59	&	0.51	&	0.58 \\
          &  Not smoked at least 100 cigarettes   & 0.41	&	0.49	&	0.42 \\
          & \\
 EXERANY2 & Exercise in past 30 days             & 0.22	&	0.40	&	0.25 \\
          & No exercise in past 30 days          & 0.78	&	0.60	&	0.75 \\
          & \\
 WEIGHT2  & Body weight in pounds            & 180.5 & 176.9    & 180.0 \\
\bottomrule
\end{longtable}
\normalsize

\doublespacing

As shown in Tables 6 and 7, there are differences between the target population and the web and non-web populations in the estimated
distributions of some of the covariates.  For example, 0.19 of the full population are age 65 or older, 0.14 of the web population are, and
0.45 of the non-web are 65+. In the target population, 0.59 are employed for wages, 0.65 are in the web population, but only 0.30 of the
non-web are. About 8\% of the web population have a grade 11 education or less while 13\% of the full population does; 33\% of the web
population attended four or more years of college while 29\% of the full population did. For the analysis variables in Table 7, 4.3\% of the
target population have ever been diagnosed with a heart attack while 3.1\% of the web population and 10.7\% of the non-web population have.
Similar differences occur for diabetes and stroke. Although the percentage point differences are small between the web and full populations,
the relative differences are substantial. For example, heart attacks in the web population are 72\% (0.031/0.043) of those in the full
population; diabetes in the web population is 80\% of the full population rate; strokes in the web population are 72\% of those in the full
population. Consequently, calibrating the matched sample may reduce bias and variance as long as the covariates in Table 6 are predictive of
the $Y$'s.  However, it is clear that weighting a sample from the web population will have to achieve a considerable amount of bias correction
in order to produce good estimates for the full, target population.

Also noteworthy are the substantial differences between the web and non-web subpopulations. The non-web people are older, more likely to be
Black and non-Hispanic, more likely to not be in the labor force, less educated, lower income, and more likely to have smoked than the web persons. The non-web
people are also much more likely to have had heart attacks, diabetes, and strokes. Our focus is on using a sample from the web population to
make estimates for the full population, but any attempt to use a sample from the web population to represent the non-web population seems
doomed to failure.  In general, a nonprobability sample that has serious coverage problems cannot be expected to produce good estimates for
poorly covered domains.

\pagebreak

\singlespacing
{\centering
Table 7~~{Proportions of the web, non-web, and total populations that have been told by a medical professional that they have three health
conditions}}\vspace{1ex}
\begin{longtable}{p{.35\linewidth}p{.20\linewidth}p{.20\linewidth}p{.20\linewidth}}
\toprule
Condition & Web & Non-web & Total pop \\
\midrule
 Heart attack (CVDINFR4) &	  0.031 & 0.107 & 0.043 \\
 Diabetes (DIABETE3)     &    0.093 & 0.233 & 0.116 \\
 Stroke (CVDSTRK3)       &    0.020 & 0.076 & 0.029 \\
\bottomrule
\end{longtable}
\doublespacing

To apply the proposed matching method, simple random samples are selected from the BRFSS web subsample and from the BRFSS full sample. Using
equal probability sampling preserves any  differences between the web and full samples and, in particular, any coverage defects in the web
sample. The size of the $S_p$ probability sample was $n=500$ while the size of the initial $S_{np}$ web sample was $M=3000$. The BRFSS raked
weights for persons in $S_p$ were adjusted to equal $\tilde{w}_j = (N/n)\ast \texttt{X\_LLCPWT}$ where $N=315,669$. Since the BRFSS design
weights did not include a nonresponse adjustment and, consequently, did not sum to an estimate of the size of the target population, we
computed a nonresponse-adjusted design weight for each person in $S_p$ as $\tilde{w}_{\pi j} = (N/n) \ast \texttt{X\_WT2RAKE} \ast f_{NR}$
where $f_{NR}$ is the sum of \texttt{X\_LLCPWT} over the sum of \texttt{X\_WT2RAKE}.

The samples, $S_p$ and $S_{np}$, are combined and the propensity of being in $S_p$ is estimated via logistic regression. The $n$ closest
matches in $S_{np}$, found using the R package \texttt{Matching}, are retained for estimation. The matching reduces the size of $S_{np}$ to be
the same ($n=500$) as that of $S_p$. The weights $\tilde{w}_j$ and $\tilde{w}_{\pi j}$ from the matching person in $S_p$ are assigned to person
$j$ in $S_{np}$.  These weights were used to calculate estimated proportions, $\widehat{\overline{Y}}_{M1}$, $\widehat{\overline{Y}}_{M2}$,
$\widehat{\overline{Y}}_{MC1}$, $\widehat{\overline{Y}}_{MC2}$, and their associated variance estimators. Estimators of the proportions of
persons who reported heart attacks, diabetes, or strokes were computed based on the estimators of totals divided by $\widehat{N} =
\sum_{S_{np}} \tilde{w}_j$. Because of the way full-sample BRFSS weights are constructed, the variation of $\widehat{N}$ from sample to sample
is minimal so that $\widehat{N}$ is treated as a constant for variance estimation.

For $\widehat{\overline{Y}}_{MC1}$ and $\widehat{\overline{Y}}_{MC2}$ the calibration model used main effects for \texttt{SEX}, \texttt{X\_AGE},
\texttt{MARITAL}, \texttt{EMPLOY1}, \texttt{EDUCA}, \texttt{INCOME2}, \texttt{EXERANY2}, and \texttt{SMOKE100} plus the continuous
variable \texttt{WEIGHT2}. After some testing, the race variable was not included since it did not improve predictions once the other
covariates were in the model. Calibration was done with the R \texttt{survey} package \citep{Lumley.2020}.

We also computed two versions of the doubly robust estimator for comparison. The two alternatives differed in the propensity model used. The
first, $\widehat{\overline{Y}}_{DR1}$, used a propensity model with the same covariates as the calibration model for
$\widehat{\overline{Y}}_{MC1}$ and $\widehat{\overline{Y}}_{MC2}$.  The second, $\widehat{\overline{Y}}_{DR2}$, used a propensity model that
included an intercept, the interactions of \texttt{INCOME2} with \texttt{X\_AGE}, \texttt{EDUCA} with
\texttt{X\_AGE}, and \texttt{INCOME2} with
\texttt{EDUCA}. These interactions were determined from a regression tree analysis, and the covariates were recoded for the interactions to be
binary.  \texttt{INCOME2} was recoded to less than or greater than or equal to \$25,000; \texttt{X\_AGE} to less than 55 years or greater than or
equal to 55 years; \texttt{EDUCA} to less than high school or high school or more. The logistic propensity model for being in $S_{np}$ based on
the merged dataset of $S_p$ and $S_{np}$ was estimated using the method described in \citet{Wang.2021}. For both doubly robust alternatives,
the same calibration model was used as for $\widehat{\overline{Y}}_{MC1}$ and $\widehat{\overline{Y}}_{MC2}$.

This process was repeated $5,000$ times for each of the three analysis variables. The relative biases, the variances and the mean squared
errors (MSEs) of the three point estimators across the $5,000$ samples are summarized in Table 8. For all three analysis variables the biases
of $\widehat{Y}_{M1}$ and $\widehat{Y}_{M2}$ are positive, ranging from 4.8\% for diabetes with $M1$ to 15.7\% for heart attack for $M2$.
Recall that $M1$ is a type of $\pi$-estimator with the $\pi$-weight taken from the matched case in the probability sample. In this example,
$M2$ is a raked estimator with the weight being the raked weight from the matched case in $S_p$. In contrast, the $MC1$, $MC2$, $DR1$, and
$DR2$ estimators have serious negative biases, ranging from -21.6\% to -17.5\%. The ordering of the MSEs varies, although $DR2$ has the
smallest MSE for two of the three analysis variables. None of the alternatives is able to correct for the undercoverage by the the web sample
of the full population.

Table 9 shows the percent relative biases of the variance estimators with respect to the empirical variance of each estimator of the proportion
and with respect to the empirical MSE. These are labeled RB.Empvar (\%) and RB.MSE (\%). For the most part, the relative biases are negative.
With respect to the MSE, all are negative owing to the biases of the point estimators of the proportions which inflate the MSEs. The coverage
rates for 95\% normal approximation confidence intervals is generally poor because the intervals are centered at the wrong place due to the
biases of the estimators of proportions. Only the combination of $\widehat{\overline{Y}}_{M1}$ with $v_{R\pi\xi}$ has coverage rates above
90\%.

\pagebreak

Finally, as an experiment we also increased the sample sizes to $n=1000$ for the nonprobability sample and $M=5000$ for the initial probability sample. The increased sample sizes had no effect on the biases of the point estimates of means. (Results are omitted here.)

\singlespacing														
{\centering														
Table 8~~{Simulation study with BRFSS population: Monte Carlo percent relative biases, variances and mean squared errors of the point
estimators}}														
\begin{longtable}{p{.18\linewidth}p{.18\linewidth}p{.18\linewidth}p{.18\linewidth}p{.18\linewidth}}														 
\toprule														
Estimator & \hfil	Relative Bias	&	\hfil	Variance	    \hfil	&	\hfil MSE \hfil & \hfil	Ratio to \\
          & \hfil	(\%)	\hfil	&	\hfil	($\times 10^4$)	\hfil	&	\hfil ($\times 10^4$)	& \hfil	 min MSE	\\
\midrule														
\textbf{Heart attack}   &       &                   &       &             \\
\hfil $\widehat{Y}_{M1}$  & \hfil	12.4	& \hfil 2.57	&	\hfil	2.85	&	\hfil	1.41	 \\
\hfil $\widehat{Y}_{M2}$  & \hfil	15.7	& \hfil 3.76	&	\hfil	4.22	&	\hfil	2.08	 \\
\hfil $\widehat{Y}_{MC1}$ & \hfil	-20.5	& \hfil	1.73	&	\hfil	2.51	&	\hfil	1.24	 \\
\hfil $\widehat{Y}_{MC2}$ & \hfil	-20.3	& \hfil	2.03	&	\hfil	2.78	&	\hfil	1.38	 \\
\hfil $\widehat{Y}_{DR1}$ & \hfil	-21.6	& \hfil	1.61	&	\hfil	2.47	&	\hfil	1.22	 \\
\hfil $\widehat{Y}_{DR2}$ & \hfil	-21.3	& \hfil	1.19	&	\hfil	2.02	&	\hfil	1.00	 \\
\textbf{Diabetes}    	&			        &		    &				 \\
\hfil $\widehat{Y}_{M1}$  & \hfil	4.8	    & \hfil	5.71	&	\hfil	6.02	&	\hfil	1.00	\\
\hfil $\widehat{Y}_{M2}$  & \hfil	6.4	    & \hfil	8.07	&	\hfil	8.62	&	\hfil	1.43	\\
\hfil $\widehat{Y}_{MC1}$ & \hfil	-20.1	& \hfil	4.47	&	\hfil	9.88	&	\hfil	1.64	\\
\hfil $\widehat{Y}_{MC2}$ & \hfil	-19.8	& \hfil	5.26	&	\hfil	10.53	&	\hfil	1.75	\\
\hfil $\widehat{Y}_{DR1}$ & \hfil	-20.6	& \hfil	3.78	&	\hfil	9.48	&	\hfil	1.58	\\
\hfil $\widehat{Y}_{DR2}$ & \hfil	-20.3	& \hfil	2.92	&	\hfil	8.46	&	\hfil	1.41	\\
\textbf{Stroke}	    &			        &			&			        \\
\hfil $\widehat{Y}_{M1}$  & \hfil	11.2	& \hfil	1.73	&	\hfil	1.83	&	\hfil	1.46	\\
\hfil $\widehat{Y}_{M2}$  & \hfil	15.5	& \hfil	2.59	&	\hfil	2.80	&	\hfil	2.24	\\
\hfil $\widehat{Y}_{MC1}$ & \hfil	-18.4	& \hfil	1.33	&	\hfil	1.62	&	\hfil	1.29	\\
\hfil $\widehat{Y}_{MC2}$ & \hfil	-17.5	& \hfil	1.59	&	\hfil	1.85	&	\hfil	1.47	\\
\hfil $\widehat{Y}_{DR1}$ & \hfil	-20.3	& \hfil	1.17	&	\hfil	1.52	&	\hfil	1.21	\\
\hfil $\widehat{Y}_{DR2}$ & \hfil	-20.1	& \hfil	0.91	&	\hfil	1.25	&	\hfil	1.00	\\
\bottomrule														
\end{longtable}														
\doublespacing														

\pagebreak
\singlespacing														
{\centering														
Table 9~~{Simulation study with BRFSS population: Percent relative biases and 95\% confidence interval coverages of the variance estimators}}														

\begin{longtable}{p{.18\linewidth}p{.18\linewidth}p{.18\linewidth}p{.18\linewidth}p{.18\linewidth}}														 
\toprule														
Estimator & \hfil	RB.Empvar (\%)	&	\hfil	RB.MSE (\%)	    \hfil	&	\hfil CI coverage (\%)\\
\midrule														
\textbf{Heart attack}   &       &        &       &             \\
\hfil $v_{\xi}(\widehat{Y}_{M1})$     & \hfil	-28.4	& \hfil	-35.5	& \hfil	89.6  \\
\hfil $v_{R\pi}(\widehat{Y}_{M1})$    & \hfil   -4.0	& \hfil	-13.4	& \hfil	92.8  \\
\hfil $v_{R\pi\xi}(\widehat{Y}_{M1})$ & \hfil	-10.1   & \hfil -18.9   & \hfil 92.6  \\
\hfil $v_{\xi}(\widehat{Y}_{M2})$     & \hfil	-26.0	& \hfil	-33.9	& \hfil	89.6  \\
\hfil $v_{R\pi}(\widehat{Y}_{M2})$    & \hfil	-22.7	& \hfil	-31.1	& \hfil	89.5  \\
\hfil $v_{R\pi\xi}(\widehat{Y}_{M2})$ & \hfil	-26.9	& \hfil	-34.8	& \hfil	89.4  \\
\hfil $v_{\xi}(\widehat{Y}_{MC1})$    & \hfil	7.5	    & \hfil	-25.8	& \hfil	81.6  \\
\hfil $v_{R}(\widehat{Y}_{MC1})$      & \hfil	-14.4	& \hfil	-40.9	& \hfil	73.1  \\
\hfil $v_{\xi}(\widehat{Y}_{MC2})$    & \hfil	-8.1	& \hfil	-33.1	& \hfil	80.4  \\
\hfil $v_{R\pi}(\widehat{Y}_{MC2})$    & \hfil	-5.0	& \hfil	-30.9	& \hfil	77.2  \\
\hfil $v_{R\pi\xi}(\widehat{Y}_{MC2})$ & \hfil	13.8	& \hfil	-17.2	& \hfil	84.4  \\
\textbf{Diabetes}							  \\
\hfil $v_{\xi}(\widehat{Y}_{M1})$     & \hfil	-26.3	& \hfil	-30.1	& \hfil	89.8	  \\
\hfil $v_{R\pi}(\widehat{Y}_{M1})$     & \hfil	3.1	    & \hfil	-2.2	& \hfil	94.8	  \\
\hfil $v_{R\pi\xi}(\widehat{Y}_{M1})$  & \hfil  2.1     & \hfil -3.1    & \hfil 94.9      \\
\hfil $v_{\xi}(\widehat{Y}_{M2})$      & \hfil	-25.0	& \hfil	-29.8	& \hfil	89.9	  \\
\hfil $v_{R\pi}(\widehat{Y}_{M2})$     & \hfil	-8.9	& \hfil	-14.8	& \hfil	92.9	  \\
\hfil $v_{R\pi\xi}(\widehat{Y}_{M2})$  & \hfil	-9.7	& \hfil	-15.5	& \hfil	92.9	  \\
\hfil $v_{\xi}(\widehat{Y}_{MC1})$     & \hfil	-5.3	& \hfil	-57.2	& \hfil	72.9	  \\
\hfil $v_{R}(\widehat{Y}_{MC1})$       & \hfil	-13.4	& \hfil	-60.9	& \hfil	68.4	  \\
\hfil $v_{\xi}(\widehat{Y}_{MC2})$     & \hfil	-19.6	& \hfil	-59.8	& \hfil	71.7	  \\
\hfil $v_{R\pi}(\widehat{Y}_{MC2})$     & \hfil	1.5	    & \hfil	-49.3	& \hfil	75.6	  \\
\hfil $v_{R\pi\xi}(\widehat{Y}_{MC2})$  & \hfil	8.3	    & \hfil	-45.9	& \hfil	79.4	  \\
\textbf{Stroke}							  \\
\hfil $v_{\xi}(\widehat{Y}_{M1})$     & \hfil	-25.6	& \hfil	-29.9	& \hfil	87.9	  \\
\hfil $v_{R\pi}(\widehat{Y}_{M1})$    & \hfil	-1.8	& \hfil	-7.5	& \hfil	90.1	  \\
\hfil $v_{R\pi\xi}(\widehat{Y}_{M1})$ & \hfil   -9.7    & \hfil -14.9   & \hfil 90.5      \\
\hfil $v_{\xi}(\widehat{Y}_{M2})$     & \hfil	-24.7	& \hfil	-30.2	& \hfil	87.8	  \\
\hfil $v_{R\pi}(\widehat{Y}_{M2})$    & \hfil	-24.6	& \hfil	-30.1	& \hfil	87.4	  \\
\hfil $v_{R\pi\xi}(\widehat{Y}_{M2})$ & \hfil	-29.8	& \hfil	-35.0	& \hfil	87.4	  \\
\hfil $v_{\xi}(\widehat{Y}_{MC1})$    & \hfil	-2.3	& \hfil	-19.8	& \hfil	79.7	  \\
\hfil $v_{R}(\widehat{Y}_{MC1})$      & \hfil	-16.1	& \hfil	-31.1	& \hfil	73.2	  \\
\hfil $v_{\xi}(\widehat{Y}_{MC2})$    & \hfil	-18.3	& \hfil	-29.8	& \hfil	78.8	  \\
\hfil $v_{R\pi}(\widehat{Y}_{MC2})$    & \hfil	-13.3	& \hfil	-25.6	& \hfil	76.9	  \\
\hfil $v_{R\pi\xi}(\widehat{Y}_{MC2})$ & \hfil	-1.7	& \hfil	-15.6	& \hfil	82.2	  \\
\bottomrule														
\end{longtable}														
\doublespacing

\section{Conclusion} \label{sec:conclusion}
In this article we present several alternative estimators when a nonprobability sample, $S_{np}$, is matched to a probability sample, $S_p$.
The general setting is that the nonprobability sample is weighted by assigning the weight from an $S_p$ unit to its matched unit in the
nonprobability sample. Particular cases are (i) the weight from $S_p$ is its $\pi$-weight, (ii) the weight from $S_p$ is a GREG weight, (iii)
case (i) with the nonprobability sample being calibrated with a linear model, and (iv) case (ii) with $S_{np}$ calibrated with a linear model.
Under some restrictive conditions, these estimators can be approximately unbiased. The key requirement is that the actual propensity of a
unit's being observed in the nonprobability sample should be equal to the inclusion probability of the unit that it is matched to in the
probability sample.

Three simulation studies illustrated several points about the matched estimator and the doubly robust estimator, which is included for
comparison. Study I used artificial data where the variable to be analyzed follows a linear model with a single covariate $X$, which was also
used to create strata. The sample designs for both $S_p$ and $S_{np}$ were stratified simple random sampling with the design for $S_{np}$
treated as unknown. In this case, matching on $X$ was reliable and all estimators were unbiased. In fact, three of four of the matching
estimators had a smaller MSE than the doubly robust estimator.

The second simulation used the same artificial population and $S_p$ sample design as Study I, but $S_{np}$ was selected with probabilities
(treated as unknown) that decreased with $X$. In this example, the inclusion probabilities for the nonprobability sample are far from those in
the probability sample used for matching. Consequently, the matched estimators without calibration are biased. However, calibration corrects
the biases and the matched, calibrated estimator has a smaller MSE than the doubly robust estimator.

The third simulation used a real population (the US Behavioral Risk Factor Surveillance Survey, BRFSS) in which persons who had accessed the
internet in the previous 30 days were treated as a nonprobability sample from the full US adult population. Since there was no control over how
the nonprobability units were selected, this mirrored a situation that would be faced in practice. The prevalence of three health conditions
was estimated. The prevalences differed considerably between the part of the population that was covered by $S_{np}$ and the part that was not.
The persons who did not use the internet were older, less educated, lower income, and less healthy than the internet users. These differences
led to all estimators in the study being biased. Calibrating the matching estimators on a list of covariates did not correct the biases. In
addition, doubly robust estimation, which has been touted as one of the better options, had substantial biases that were larger then those of
the best matching estimators.

The failure in the real data study has several, potential contributing factors, including poor matches between the nonprobability and
probability units, inadequate models for the propensity of being observed in the nonprobability sample, and poor calibration models for
predicting the health characteristics analyzed. However, the facts that the nonprobability sample does not cover the target population, and the
noncovered units differ both on the distributions of the analytic variables and covariates is the critical problem. Some diagnostics have been
devised for detecting non-ignorability of selection of a nonprobability sample \citep[e.g., see][]{Andridge.2019, Little.2019}. These
diagnostics will signal non-ignorability if the means of covariates in $S_{np}$ and the target population are sufficiently different. Thus,
they might be a way forward in the BRFSS application.

However, if the variables to be analyzed differ between $S_{np}$ and the target
population but covariate distributions do not, the diagnostics will not alert an analyst to trouble, and poor inferences will still be made from the nonprobability sample. The type of coverage error in the BRFSS study is an example of what can happen in nonprobability samples, generally, and may be a problem that no amount of sophisticated mathematics is likely to correct.

\appendix

\section{Appendix}
This appendix shows the details of variance calculations given in earlier sections.  Several assumptions are used in the results below. These
apply as $N$ and $n \rightarrow \infty$.
\begin{enumerate}[(i)]
    \item $\pi_j = O(N/n)$, $R_j = O(N/n)$ and $n/N \rightarrow 0$ \label{itm:pi.ord}
    \item $\tilde{\mathbf{A}}_U$ and $\mathbf{A}_U^{\ast}$ are $O(N)$       \label{itm:AU.ord}
    \item $V_{\pi}\left( \mathbf{X}_p \right) = O(N^2/n)$   \label{itm:Vpi.ord}
    \item $V_{R}\left( \mathbf{X}_{np} \right) = O(N^2/n)$  \label{itm:VR.ord}
    \item When $R_j = \pi_j$, $N^{-1}\tilde{\mathbf{A}}_{p}$ and  $N^{-1}\tilde{\mathbf{A}}_{np}(\pi)$ both converge in probability to \\
        $N^{-1}\tilde{\mathbf{A}}_{U} = N^{-1}\sum_U \mathbf{x}_j \mathbf{x}_j^T/\tilde{\sigma}_j^2$.            \label{itm:ApAnp.conv}
    \item $N^{-1}\mathbf{A}_{np}^{\ast}(\tilde{w})$ converges in probability to $N^{-1}\tilde{\mathbf{A}}_{U}^{\ast} = N^{-1}\sum_U
        \mathbf{x}_j \mathbf{x}_j^T/\tilde{\sigma}_j^{\ast}$.            \label{itm:Atildenp.conv}
    \item When $R_j = \pi_j$, $\widetilde{\mathbf{A}}_p^{-1} \sum_{S_{np}} \frac{\mathbf{x}_j y_j}{\pi_j \widetilde{\sigma}_j^2}
        \overset{p}{\to} \widetilde{\mathbf{B}}_U$ and $\left[\widetilde{\mathbf{A}}_{np}^{\ast}(\widetilde{w})\right]^{-1} \sum_{S_{np}}
        \frac{\mathbf{x}_j y_j}{\pi_j \sigma_j^{\ast 2}} \overset{p}{\to} \mathbf{B}_U^{\ast}$
        \label{itm:Bconv}
    \item When $R_j = \pi_j$, $\sqrt{n}\left(\widehat{\mathbf{X}}_p - \mathbf{X}_U \right)/N$, $\sqrt{n}\left(\widehat{\mathbf{X}}_{np}(\pi)
        - \mathbf{X}_U \right)/N$, and $\sqrt{n}\left(\widehat{\mathbf{X}}_{np}(\widetilde{w}) - \mathbf{X}_U \right)/N$ are asymptotically
        multivariate normal with mean $\mathbf{0}$.  \label{itm:ANorm}
\end{enumerate}

\subsection{$\xi$-expectation of the With-replacement Variance Estimator under Case (i)} \label{app:vRexp}

To compute the $\xi$-expectation of $v_R\left(\widehat{Y}_{M1} \right)$ in section \ref{subsec:Vcase.i} under case (i), define $r_j =
\widetilde{w}_{j}y_{j} - \frac{1}{n} \sum_{j^{\prime} \in S_{np}} \widetilde{w}_{j^\prime}y_{j\prime}$. Since $\widetilde{w}_j = \pi_j^{-1}$,
this can be rewritten as
\[r_j = \frac{n-1}{n} \frac{y_j}{\pi_j} - \frac{1}{n} \sum_{j^{\prime} \neq j \in S_{np}} \frac{y_{j^\prime}}{\pi_{j^\prime}}. \]
The $\xi$-expectation of $r_j^2$ is then
\begin{align*}
    E_{\xi}(r_j^2) &= V_{\xi}(r_j) + [ E_{\xi}(r_j)]^2 \\
    &= \left(\frac{n-1}{n} \right)^2 \frac{\sigma_j^2}{\pi_j^2} + \frac{1}{n^2} \left( \sum_{j^\prime \neq j \in S_{np}}
    \frac{\sigma_{j^\prime}^2}{\pi_{j^\prime}^2} \right) + \Bigg\{\left( \frac{\mathbf{x}_j}{\pi_j} - \frac{1}{n}\sum_{{\color{green}j^{\prime}
    \in S_{np}}}\frac{\mathbf{x}_{j^{\prime}}}{\pi_{j^{\prime}}} \right)^T \boldsymbol{\beta} \Bigg\}^2\,.
\end{align*}
Adding and subtracting $\sigma_j^2/\pi_j^2$ in the second term, summing over $S_{np}$, and doing some algebra leads to
\[ E_{\xi}\left(v_{R\pi} \right) = \sum_{S_{np}} \frac{\sigma_j^2}{\pi_j^2} + \frac{n}{n-1} \sum_{j \in S_{np}} \Bigg\{\left(
\frac{\mathbf{x}_j}{\pi_j} - \frac{1}{n}\sum_{j^{\prime} \in S_{np}}\frac{\mathbf{x}_{j^{\prime}}}{\pi_{j^{\prime}}} \right)^T
\boldsymbol{\beta} \Bigg\}^2 \]
as noted in section \ref{subsec:Vcase.i}. That is, $v_{R\pi}$ is an overestimate of the model variance under \eqref{eq:model}. However, because
$\widehat{Y}_{M1}$ is model-biased, $v_{R\pi}$ will not appropriately estimate the $\xi$ mean square error despite its overestimating the
$\xi$-variance.

To derive the $R\pi\xi$-variance, note that
\begin{align*}
V_{R\pi\xi}\left( \widehat{Y}_{M1} \mid S_p,S_{np}\right) &= V_{R\xi}\left( \widehat{Y}_{M1} \mid S_p,S_{np}\right) \\
    &= E_R \left\{V_{\xi}\left( \widehat{Y}_{M1} \mid S_p,S_{np}\right) \right\} +
        V_R \left\{E_{\xi}\left( \widehat{Y}_{M1} \mid S_p,S_{np} \right) \right\}.
\end{align*}
Using the independence of the $Y$'s under \eqref{eq:model}, the first term is $\sum_U \sigma_j^2 / \pi_j$.  The second term is $V_R
\left\{E_{\xi}\left( \widehat{Y}_{M1} \mid S_p,S_{np} \right) \right\} = V_R\left(\widehat{\mathbf{X}}_{np}(\pi)^T \boldsymbol{\beta}\right) =
\boldsymbol{\beta}^T V_R\left(\widehat{\mathbf{X}}_{np}(\pi) \right) \boldsymbol{\beta}$. Combining gives the expression shown in
\eqref{eq:YM1.VRpi}.

\subsection{Variance of Matched Estimator $\widehat{Y}_{M2}$ under Case (ii)} \label{app:YM2c2var}
Following similar steps to those in \citet[sec.6.6]{Sarndal1992} and using condition  \ref{itm:Bconv}, $\widehat{Y}_{M2}$ can be approximated
as
\begin{equation} \label{eq:appYM2approx}
    \widehat{Y}_{M2} \doteq \widehat{Y}_{np}(\pi) + \left(\mathbf{X}_U - \widehat{\mathbf{X}}_p  \right) \tilde{\mathbf{B}}_U\,,
\end{equation}
where $\tilde{\mathbf{B}}_U = \left(\sum_U \frac{\mathbf{x}_j \mathbf{x}_j^T} {\pi_j \tilde{\sigma}_j^2} \right)^{-1} \left(\sum_U
\frac{\mathbf{x}_j y_j} {\pi_j \tilde{\sigma}_j^2} \right)$.

Using the formula for total variance across the $R$ and $\pi$ distributions (denoted by $V_{R\pi}$) gives
\begin{equation} \label{eq:totvarRpi}
    V_{R\pi}(\widehat{Y}_{M2}) = E_{R} V_{\pi}(\widehat{Y}_{M2} \mid S_{np}) + V_R E_{\pi}(\widehat{Y}_{M2} \mid S_{np})\,.
\end{equation}
Working term by term in \eqref{eq:totvarRpi} and using the approximation to $\widehat{Y}_{M2}$ in equation \eqref{eq:appYM2approx}, {\color{green}we have}
\[E_R V_{\pi}(\widehat{Y}_{M2} \mid S_{np}) \doteq E_R V_{\pi} \left( \widehat{Y}_{np}(\pi) + \left( \mathbf{X}_U - \widehat{\mathbf{X}}_{p}
\right)^T \widetilde{\mathbf{B}}_U \bigg\vert S_{np} \right)= \widetilde{\mathbf{B}}_U^T V_{\pi}\left(\widehat{\mathbf{X}}_{p} \right)
\widetilde{\mathbf{B}}_U \,,\]
because $\widehat{Y}_{np}(\pi)$ has zero $R$-variance given that $S_{np}$ is fixed. To get the second term in \eqref{eq:totvarRpi}, note that
$V_R E_{\pi}(\widehat{Y}_{M2} \mid S_{np}) \doteq V_R\left(\widehat{Y}_{np}(\pi) \right)$ assuming that $\widehat{\mathbf{X}}_{p}$ is
$\pi$-unbiased. Combining {\color{green}these} results, the variance across the $R$- and $\pi$-distributions is
\begin{equation*}
    V_{R\pi}\left( \widehat{Y}_{M2} \right) \doteq  V_R \left( \widehat{Y}_{np}(\pi) \right) + \widetilde{\mathbf{B}}_U^T
    V_{\pi}\left(\widehat{\mathbf{X}}_p \right) \widetilde{\mathbf{B}}_U.
\end{equation*}
as shown in \eqref{eq:VRpi.M2c2}.

Turning to the $R\pi\xi$-variance, the total variance formula is given by \eqref{eq:3partvar}. The $E_{\pi}V_{\xi}$ term is $\sum_{S_{np}}
\tilde{\sigma}_j^2 E_{\pi}(g_j^2) / \pi_j^2$. Using a Taylor series approximation as in \citet[sec.6.6]{Sarndal1992}, we have
\begin{equation} \label{eq:gapprox}
   g_j \doteq \pi_j^{-1}\left[1 + \left(\mathbf{X}_U - \widehat{\mathbf{X}}_p \right)^T \tilde{\mathbf{A}}_U^{-1}
   \mathbf{x}_j/\tilde{\sigma}_j^2 \,.\right]
\end{equation}
It follows that
\begin{align}
    E_{\pi}V_{\xi}\left(\widehat{Y}_{M2} \right) &\doteq \sum_{S_{np}} \frac{\sigma_j^2}{\pi_j^2} \left\{ 1 +
    \frac{\mathbf{x}_j^T}{\tilde{\sigma}_j^2} \tilde{\mathbf{A}}_U^{-1} E_{\pi}\left[\left(\mathbf{X}_U - \widehat{\mathbf{X}}_p\right)
    \left(\mathbf{X}_U - \widehat{\mathbf{X}}_p \right)^T \right] \tilde{\mathbf{A}}_U^{-1} \frac{\mathbf{x}_j} {\tilde{\sigma}_j^2} \right\}
    \notag \\
    &= \sum_{S_{np}} \frac{\sigma_j^2}{\pi_j^2} \left\{ 1 + \frac{\mathbf{x}_j^T}{\tilde{\sigma}_j^2} \tilde{\mathbf{A}}_U^{-1}
    V_{\pi}\left(\widehat{\mathbf{X}}_p\right) \tilde{\mathbf{A}}_U^{-1} \frac{\mathbf{x}_j} {\tilde{\sigma}_j^2} \right\}\,.
\end{align}
Thus,
\[
    E_R E_{\pi}V_{\xi}\left(\widehat{Y}_{M2} \right) = \sum_U R_j \frac{\sigma_j^2}{\pi_j^2} + \sum_U R_j \frac{\sigma_j^2}{\pi_j^2}
    \frac{\mathbf{x}_j^T}{\tilde{\sigma}_j^2} \tilde{\mathbf{A}}_U^{-1} V_{\pi}\left(\widehat{\mathbf{X}}_p\right) \tilde{\mathbf{A}}_U^{-1}
    \frac{\mathbf{x}_j} {\tilde{\sigma}_j^2}\,.
\]
Under the order assumptions at the beginning of this appendix, the first term above is $O(N^2 / n)$ while the second is $O(N^2 / n^2)$.  Thus,
we use the approximation $E_R E_{\pi}V_{\xi}\left(\widehat{Y}_{M2} \right) \doteq \sum_U R_j \sigma_j^2 / \pi_j^2$.

The second term in \eqref{eq:3partvar} is $E_R V_{\pi} E_{\xi}\left(\widehat{Y}_{M2} \right)$. Expanding and collecting terms gives
\begin{align*}
    V_{\pi} E_{\xi}\left(\widehat{Y}_{M2} \right) &= V_{\pi}\left(\sum_{S_{np}} \frac{g_j}{\pi_j} \mathbf{x}_j \boldsymbol{\beta} \right) \\
    &= V_{\pi}\left( \sum_{S_{np}} \frac{\mathbf{x}_j}{\pi_j} \boldsymbol{\beta} + \left(\mathbf{X}_U - \mathbf{X}_p \right)^T
    \tilde{\mathbf{A}}_p^{-1} \tilde{\mathbf{A}}_{np} \boldsymbol{\beta} \right)\,.
\end{align*}
Under condition (v) above, $\tilde{\mathbf{A}}_p^{-1} \tilde{\mathbf{A}}_{np}$ converges to the $C \times C$ identity matrix and $E_R V_{\pi}
E_{\xi}\left(\widehat{Y}_{M2} \right) = \boldsymbol{\beta}^T V_{\pi}\left(\widehat{\mathbf{X}}_p \right) \boldsymbol{\beta}$.

The third term in \eqref{eq:3partvar} is $V_R E_{\pi} E_{\xi}\left(\widehat{Y}_{M2} \right)$. First, compute $E_{\pi}
E_{\xi}\left(\widehat{Y}_{M2} \right) = E_{\pi}\left(\sum_{S_{np}} \frac{g_j}{\pi_j}\mathbf{x}_j \boldsymbol{\beta} \right)$. Using the
approximation to $g_j$ in \eqref{eq:gapprox}, $E_{\pi}(g_j) \doteq 1$ and $E_{\pi} E_{\xi}\left(\widehat{Y}_{M2} \right) \doteq
\widehat{\mathbf{X}}_{np}(\pi) \boldsymbol{\beta}$. Consequently, the third term is $V_R E_{\pi} E_{\xi}\left(\widehat{Y}_{M2} \right) \doteq
\boldsymbol{\beta}^T V_R\left(\widehat{\mathbf{X}}_{np}(\pi)\right) \boldsymbol{\beta}$. Combining results for the three terms in
\eqref{eq:3partvar} gives
\[
    V_{R\pi\xi}\left(\widehat{Y}_{M2} \right) \doteq \sum_U R_j \frac{\sigma_j^2}{\pi_j^2} + \boldsymbol{\beta}^T
    V_{\pi}\left(\widehat{\mathbf{X}}_p \right) \boldsymbol{\beta} + \boldsymbol{\beta}^T V_R\left(\widehat{\mathbf{X}}_{np}(\pi)\right)
    \boldsymbol{\beta}
\]
as shown in \eqref{eq:VRpixi.M2c2}.

\subsection{Approximation to $\widehat{Y}_{MC2}$ in case (ii)} \label{app:YMC2approx}
When $S_p$ has case (ii) weights, $\widetilde{w}_j = g_j/\pi_j$ with $g_j$ defined in \eqref{eq:gj}. The matched estimator after calibration
then equals
\[ \widehat{Y}_{MC2} = \sum_{S_{np}} g_j^{\ast} g_j y_j/ \pi_j\,,
\]
where
\[ g_j^{\ast} = 1 + \left(\mathbf{X}_U - \widehat{\mathbf{X}}_{np}(\widetilde{w})\right)^T
\left[\widetilde{\mathbf{A}}_{np}^{\ast}(\widetilde{w})\right]^{-1} \mathbf{x}_j/\sigma_j^{\ast 2}\,.
\]
Multiplying $g_j^{\ast}$ by $g_j$ defined in \eqref{eq:gj} and substituting in the formula for $\widehat{Y}_{MC2}$ gives
\begin{align} \label{eq:YMC2expand}
    \widehat{Y}_{MC2} &= \widehat{Y}_{np}(\pi) \; + \;
        \left(\mathbf{X}_U - \widehat{\mathbf{X}}_p\right)^T \widetilde{\mathbf{A}}_p^{-1} \sum_{S_{np}} \frac{\mathbf{x}_j y_j}{\pi_j
        \widetilde{\sigma}_j^2} \; + \;
        \left(\mathbf{X}_U - \widehat{\mathbf{X}}_{np}(\widetilde{w})\right)^T
        \left[\widetilde{\mathbf{A}}_{np}^{\ast}(\widetilde{w})\right]^{-1} \sum_{S_{np}} \frac{\mathbf{x}_j y_j}{\pi_j \sigma_j^{\ast 2}}
        \notag \\
        &+ \left(\mathbf{X}_U - \widehat{\mathbf{X}}_p\right)^T \widetilde{\mathbf{A}}_p^{-1} \sum_{S_{np}} \frac{\mathbf{x}_j \mathbf{x}_j^T
        y_j}{\pi_j \widetilde{\sigma}_j^2 \sigma_j^{\ast 2}}  \left[\widetilde{\mathbf{A}}_{np}^{\ast}(\widetilde{w})\right]^{-1} \left(
        \mathbf{X}_U - \widehat{\mathbf{X}}_{np}(\widetilde{w})\right).
\end{align}
Using conditions \ref{itm:ApAnp.conv}, \ref{itm:Bconv}, and \ref{itm:ANorm}, the orders of the second, third, and fourth terms in
\eqref{eq:YMC2expand} are $O_p(N/\sqrt{n})$, $O_p(N/\sqrt{n})$, and $O_p(N/n)$.  The calibrated estimator can then be approximated by
\begin{equation} \label{eq:YMC2approx}
    \widehat{Y}_{MC2} \doteq \widehat{Y}_{np}(\pi) \; + \;
        \left(\mathbf{X}_U - \widehat{\mathbf{X}}_p\right)^T \widetilde{\mathbf{B}}_U \; + \;
        \left(\mathbf{X}_U - \widehat{\mathbf{X}}_{np}(\widetilde{w})\right)^T \mathbf{B}_{U}^{\ast}\,.
\end{equation}

\subsection{Variance of Matched Estimator $\widehat{Y}_{MC2}$ in case (ii)} \label{app:YMC2var}
To compute the $\xi$ model variance under case (ii), we break $\sum_U \mathbf{x}_j y_j/\widetilde{\sigma}_j^2$ and $\sum_U \mathbf{x}_j
y_j/\sigma_j^{\ast 2}$ into sums over $S_{np}$ and $U-S_{np}$. Equation \eqref{eq:YMC2approx} can then be expressed as
\begin{align*}
    \widehat{Y}_{MC2} &\doteq \sum_{S_{np}} y_j \left( \frac{1}{\pi_j} +  F_j \right)
    + \sum_{U-S_{np}} y_j F_j\,,
\end{align*}
where
\[
    F_j = \left(\mathbf{X}_U - \widehat{\mathbf{X}}_p\right)^T \widetilde{\mathbf{A}}_U^{-1} \frac{\mathbf{x}_j}{\widetilde{\sigma}_j^2} +
    \left(\mathbf{X}_U - \widehat{\mathbf{X}}_{np}(\widetilde{w})\right)^T \widetilde{\mathbf{A}}_U^{\ast -1}
    \frac{\mathbf{x}_j}{\sigma_j^{\ast 2}}\,.
\]
Applying conditions \ref{itm:AU.ord} and \ref{itm:ANorm}, $F_j=O_p\left(n^{-1/2} \right)$. Since units in $S_{np}$ and $U-S_{np}$ are
independent under model \eqref{eq:model}, the $\xi$-variance  is
\begin{align}
    V_{\xi}\left( \widehat{Y}_{MC2}\right) &\doteq \sum_{S_{np}} \sigma_j^2 \left( \frac{1}{\pi_j} +  F_j \right)^2 + \sum_{U-S_{np}}
    \sigma_j^2 F_j^2 \\
    &= \sum_{S_{np}} \left(\frac{\sigma_j^2}{\pi_j}\right)^2 \left[1 + O_p\left(N/n^{3/2}\right) \right]\,.
\end{align}

The $R\pi\xi$-variance can be calculated using the total variance formula in \eqref{eq:3partvar}. First, when $R_j=\pi_j$, $E_R E_{\pi}
V_{\xi}\left(\widehat{Y}_{MC2} \mid S_p,S_{np} \right) = \sum_U \left(\sigma_j^2 / \pi_j \right)$. The second term in \eqref{eq:3partvar} {\color{green}is}
\begin{align*}
    E_R V_{\pi} E_{\xi}\left(\widehat{Y}_{MC2} \right)
    &\doteq E_R V_{\pi}\left[\widehat{\mathbf{X}}_{np}(\pi)^T \boldsymbol{\beta} +
    \left(\mathbf{X}_U - \widehat{\mathbf{X}}_p \right)^T \boldsymbol{\beta} +
    \left(\mathbf{X}_U - \widehat{\mathbf{X}}_{np}(\widetilde{w}) \right)^T \boldsymbol{\beta}
    \right] \\
    &= \boldsymbol{\beta}^T V_{\pi}\left(\widehat{\mathbf{X}}_p \right) \boldsymbol{\beta}\,.
\end{align*}
The third term in \eqref{eq:3partvar} is
\begin{align*}
    V_R \left[E_{\pi}E_{\xi}\left(\widehat{Y}_{MC2} \right) \right] &=
    V_R E_{\pi}\left[\widehat{\mathbf{X}}_{np}(\pi)^T \boldsymbol{\beta} +
    \left(\mathbf{X}_U - \widehat{\mathbf{X}}_p \right)^T \boldsymbol{\beta} +
    \left(\mathbf{X}_U - \widehat{\mathbf{X}}_{np}(\widetilde{w}) \right)^T \boldsymbol{\beta}
    \right] \\
    &= V_R \left\{E_{\pi}\left[\widehat{\mathbf{X}}_{np}(\pi) - \widehat{\mathbf{X}}_{np}(\widetilde{w}) \right]^T \boldsymbol{\beta} \right\}\,.
\end{align*}
Rewriting the term in brackets above leads to
\begin{align*}
    \widehat{\mathbf{X}}_{np}(\pi) - \widehat{\mathbf{X}}_{np}(\widetilde{w}) &=
    \sum_{S_{np}} \frac{(1-g_j)\mathbf{x}_j}{\pi_j} \\
    &= \left( \widehat{\mathbf{X}}_p - \mathbf{X}_U \right)^T \widetilde{\mathbf{A}}_p^{-1} \sum_{S_{np}}\frac{\mathbf{x}_j
    \mathbf{x}_j^T}{\pi_j \widetilde{\sigma}_j^2} \\
    &= \left( \widehat{\mathbf{X}}_p - \mathbf{X}_U \right)^T \widetilde{\mathbf{A}}_p^{-1}\widetilde{\mathbf{A}}_{np}(\pi)\,.
\end{align*}
Applying condition \ref{itm:ApAnp.conv} implies that $V_R \left[E_{\pi}E_{\xi}\left(\widehat{Y}_{MC2} \right) \right] \doteq 0$.  Combining
results for the three terms in \eqref{eq:3partvar} yields
\begin{equation*}
    V_{R\pi\xi}\left(\widehat{Y}_{MC2} \right) \doteq \sum_U \frac{\sigma_j^2}{\pi_j} + \boldsymbol{\beta}^T
    V_{\pi}\left(\widehat{\mathbf{X}}_p \right) \boldsymbol{\beta}.
\end{equation*}
An estimator of this variance is
\[
    v_{R\pi\xi}\left(\widehat{Y}_{MC2} \right) \doteq \sum_{S_{np}} \left(\frac{\widehat{e}_j^{\ast}}{\pi_j} \right)^2 +
    \widehat{\widetilde{B}}_{np}(\pi)^T v_{\pi}\left(\widehat{\mathbf{X}}_p \right) \widehat{\widetilde{B}}_{np}(\pi)
\]
as shown in \eqref{eq:vRpixi.YMC2}.

\section*{}
\bibliographystyle{apalike}
\bibliography{References}

\end{document}